# Understanding Pedagogical Content Knowledge of Introductory Data Science Instructors: An Inaugural Framework

Sinem Demirci[1], Mine Doğucu[2], Andrew Zieffler[3], Joshua M. Rosenberg[4]

*[1]California Polytechnic State University*

*[2]University of California, Irvine*

*[3]University of Minnesota*

*[4]University of Tennessee, Knoxville, TN*

Phone: +1 805 756-0683

Email: sdemirci@calpoly.edu

As data science emerges as a distinct academic discipline, introductory data science (IDS) courses play a key role in shaping students' foundational understanding. Often taught by instructors without formal training in data science or pedagogy, these courses present a unique and globally relevant context for examining pedagogical content knowledge (PCK). Drawing on semi-structured interviews with 14 IDS instructors and their course syllabi, this study explores how IDS instructors describe and make sense of their teaching practices, which are analyzed through the lens of PCK. The findings highlight key components of PCK about IDS and offer insights into supporting instructor development. This work contributes to expanding the scope of PCK research into new interdisciplinary domains and ongoing global efforts to build capacity in data science education. It could serve as a starting point for developing a PCK framework specific to IDS.

## Introduction

Over the last few decades, data science has emerged as a multidisciplinary discipline with foundations in mathematics, statistics, computer science, and other scientific domains (Donoho, 2017; National Academies of Sciences, Engineering, and Medicine, 2018). It has gained importance in different industries, making "data scientist" one of the fastest growing jobs (The United States Bureau of Labor Statistics, 2024). Data scientist positions are on the rise in several other countries as well including but not limited to Spain, India, and Brazil (LinkedIn, 2024).

Data science as a domain did not emerge from a blank slate. Donoho (2017) summarized past work that envisioned a domain that was broader than Statistics–one that included not only theories of statistical inference but also processes of collecting or generating data; storing data; visualizing data; and communicating what one learned from the data with a range of stakeholders. As Donoho wrote, this vision is not wholly new: Tukey (1962) anticipated the changes the broader scope of data science (relative to Statistics) encompassed, and Cleveland (2001)–writing about the need for data scientists to engage with how their subject area is taught and learned–anticipated data science education.

As the demand for data scientists continues to grow, training them has become an important task for educators. Data science education has attracted attention at the school level (Heinemann et al., 2018; Rosenberg et al., 2022), the university level (Baumer, 2015; Dogucu et al., 2025), and in workplaces (Kross & Guo, 2019; McKinsey & Company, 2019). Additionally, data science has been the focus of national educational initiatives (e.g, the United Kingdom; Office for National Statistics, 2017).
Given the relatively recent emergence of data science as a discipline, there is an evident shortage of instructors with formal expertise, with demand for qualified instructors far exceeding the current capacity. Thus, these courses are primarily taught by instructors from data science related areas or industry professionals who learn the scope of their job onsite from the experts (Kross & Guo, 2019). Furthermore, the lack of consensus on both data science programs and the content and delivery of Introductory Data Science (IDS) courses (Hazzan & Mike, 2023) creates a multiplier effect, underscoring the importance of thoughtful course design. These considerations highlight the need to better understand the knowledge instructors draw upon teaching IDS effectively. With this in mind, this study aims to explore the pedagogical content knowledge (PCK) of IDS instructors, as conceptualized through the theoretical framework outlined below.

**Theoretical Framework of This Study**

To examine the teacher knowledge underpinning IDS instruction, this PCK as a theoretical lens. Coined by Shulman (1986, 1987), PCK is a distinct domain of teachers' professional knowledge that constitutes an integration of subject matter, pedagogical, and contextual knowledge (Shulman, 1987; Magnusson et al., 1999; Park & Oliver, 2008). It represents a category of knowledge that differentiates teachers from content experts (Park & Chan, 2025; Shulman, 1986; 1987). PCK has several components and they are often interrelated rather than being mutually exclusive (Hume et al., 2019).

Among the various frameworks conceptualizing PCK and its components, this study adopts Park and Oliver's (2008) PCK model for science teaching as data science intersects with multiple scientific disciplines (Mike et al., 2023). The model includes six components: (1) Orientations to Teaching Science; (2) Knowledge of Students' Understanding in Science; (3) Knowledge of Science Curriculum; (4) Knowledge of Instructional Strategies and Representations for Teaching Science; (5) Knowledge of Assessment of Science Learning; and (6) Teacher Efficacy. These six components interact continuously, influenced by their specific context. As teachers reflect and refine their PCK, the connections between these components become more cohesive, strengthening their integration (Park and Oliver, 2008).

Park and Oliver's model builds on component-based conceptualizations of PCK proposed in earlier work (e.g., Shulman, 1986; Magnusson et al., 1999) and organizes them in a hexagonal structure that highlights the dynamic interactions among components. As this model explicitly incorporates teacher efficacy, it provides an avenue for examining instructors' knowledge in IDS settings characterized by limited formal preparation and curricular uncertainty.

While this framework provides a lens for examining instructors' knowledge, it is also important to consider how IDS courses are currently structured and taught in practice. The next section reviews the focus of extant IDS courses.

**The Focus of Extant IDS Courses**

At the undergraduate level, literature on data science education spans institutional case studies and broader curricular frameworks. Farahi and Stroud (2018) documented the structure and impact of the co-curricular Michigan Data Science Team, a student-led initiative focused on data science-related outreach to community-based organizations, while Hassan and Liu (2019) proposed a model for embedding data science into existing computer science curricula. Regarding specific courses, Tucker et al. (2023) conducted an empirical analysis of student outcomes in an introductory statistics course using R (R is one of the most commonly used languages in data science courses; Schwab-McCoy et al., 2021), finding that students' initially negative views became more positive over time. There are also many data science courses taught in the context of specific domains, like *information sciences* (Hagen, 2020), though such discipline-specific courses are beyond our focus. Additionally, there have been efforts to set guidelines through expert groups and committee reports (Danyluk et al., 2021; De Veaux et al., 2017; Demchenko et al., 2021). Despite these variations, a consensus on the contents— learning objectives, topical foci, as well as key activities and assessments—of the undergraduate data science curriculum has not yet been reached.

IDS courses have also drawn attention to their role in providing foundational knowledge of data science to students. IDS courses not only help students transition to higher education but also expose students to the field, often for the first time. Thus, it is crucial for the data science education community to recognize the elements of effective IDS instruction and the needs of the educators. This need motivates a shift from identifying challenges in IDS education to examining the knowledge base that underpins effective instruction, particularly knowledge that integrates content and pedagogy in discipline-specific ways.

**From Challenge to Inquiry: The Significance of Exploring PCK about IDS**

Studying PCK for IDS instructors presents a set of challenges. One major challenge is that there is a lack of consensus in data science curricula, even in IDS, making it challenging to identify the content knowledge instructors need (Hazzan & Mike, 2023; Rosenberg & Jones, 2024). Moreover, the multidisciplinary backgrounds of instructors who teach data science and researchers and curriculum developers who create curricula further complicates any content alignment as the data science content taught varies widely across different disciplines and departments (Rosenberg & Jones, 2024). Lastly, the rapid evolution of methodologies and applications of data science further exacerbates understanding the content knowledge instructors need as the knowledge seems almost to change daily, especially in an era of AI (Artificial Intelligence) (Meng, 2025).

While these may make studying PCK for IDS instructors challenging, it also makes it a fertile ground for exploration. In PCK studies, it is argued that developing PCK is a dynamic process; although teaching experience is important, it is not sufficient on its own to enhance it (Grossman, 1990; Van Driel et al., 1998; Abell, 2008; Friedrichsen et al., 2009). Many data science instructors are either computational researchers in academia or practitioners from industry (Kross & Guo, 2019). This difference in instructor background may be advantageous for studying IDS instructors' PCK because it can provide an opportunity to examine the *formal knowledge* (Fenstermacher, 1994) and *practical knowledge* (Fenstermacher, 1994; Van Driel et al., 1998) they bring to IDS courses and how these forms of knowledge are associated with the development of their PCK.

At the same time, it is also important to recognize that the nature of PCK itself is dynamic and complex, making it difficult to find a single and definitive framework that conceptualizes and analyzes its components and development. (Barnett, 2003; Chan & Hume, 2019). Various frameworks exist (Depaepe et al., 2013; Park & Oliver, 2008; Zepke, 2013), and discrepancies among PCK studies are not uncommon (Hashweh, 2013; Settlage, 2013). Nevertheless, PCK is widely recognized as one of the teacher knowledge, transforming an instructor's pedagogical and content expertise to improve teaching and support student learning (Shulman, 1987; Han-Tosunoglu & Lederman, 2021; Park & Chan, 2025). Additionally, recent arguments (e.g., Groth, 2024) call for broadening the scope of PCK research by expanding it to other professions including those advancing quantitative reasoning, with the goal of improving public understanding and addressing urgent societal challenges. Thus, we posit that examining the nature of PCK within data science education could yield critical insights for enhancing data science education. Employing existing PCK frameworks as a starting point to explore the possible dimensions of IDS PCK as a category of teacher knowledge may also facilitate the development of a PCK framework tailored to the unique demands of data science education.

Although the teaching materials and examples of best practices in IDS courses in the literature provide some clues (e.g., Baumer, 2015; Çetinkaya-Rundel & Ellison, 2021; Schwab-McCoy et al., 2021; Yan & Davis, 2019), it remains important from a PCK perspective to examine IDS instructors' PCK holistically across components, as making these components visible can contribute to capacity building for IDS educators. In-service training and professional development initiatives could be designed to foster collective PCK within IDS education. Thus, the research question that guides this study was: *What is the nature of the Pedagogical Content Knowledge of instructors for Introductory Data Science courses?*

## Literature Review

This section reviews the literature relevant to understanding PCK in the context of IDS. First, we provide a brief overview of PCK as a conceptual lens. We then review research on PCK in data science education and related fields, highlighting existing work and identifying gaps that motivate this study.

### *Pedagogical Content Knowledge: Essentials at a Glance*

The term PCK was coined by Shulman (1986, 1987) as a category of teacher knowledge; knowledge *for* teaching that differentiates teachers from content experts. Since then, PCK frameworks have served as theoretical lenses through which researchers and educators attempt to understand components of high-quality science teaching (Gess-Newsome et al., 2019). These efforts culminated in the PCK Summits of 2012 and 2016, where leading science PCK researchers gathered to exchange ideas and work toward a 'Consensus Model' (Chan & Hume, 2019). In this study, we draw on Park and Oliver's (2008) model (see Theoretical Framework section). For an extensive review of PCK frameworks and their development, see Hume et al. (2019).

Veal and MaKinster (1999) introduced a PCK taxonomy to categorize the levels of specificity in PCK. Of the three levels—general PCK, domain-specific PCK, and topic-specific PCK—we focused on understanding the nature of *domain-specific PCK* among IDS instructors. Unlike general PCK, domain-specific PCK refers to possessing a PCK peculiar to a domain. We chose to explore IDS instructors' domain-specific PCK because IDS courses often serve as an entry point to the field, introducing students to the core topics and competencies that define data science (Gasiewski et al., 2012).

Grossman (1990) conceptualized PCK as transformed knowledge for teaching, derived from instructor's K-16 science learning experiences, teacher education programs, and teaching practice. However, if an instructor does not have a degree in education, their PCK tends to develop from their own teaching and learning experiences (Abell, 2008; Hume et al., 2019) as well as professional development activities and community engagement (Borko, 2004; Van Driel et al., 1998). This has also been a key topic in PCK research, particularly the question of how PCK evolves over time when formal teacher education programs are absent (Friedrichsen et al., 2009). For instance, Friedrichsen et al. (2009) examined participants in an Alternative Certification Program, a non-traditional route to teacher licensure, and found that prior teaching experience alone did not foster the development of robust PCK in the absence of structured pedagogical training.

Although such findings have been primarily situated in K–12 contexts, they hold important implications for higher education, where instructors are also rarely required to engage in formal pedagogical preparation (Brownell & Tanner, 2012). A recent systematic scoping review of PCK research in higher education confirmed that while interest in the topic is growing, studies remain fragmented across disciplines, and many university instructors continue to develop PCK without the support of structured professional development programs (Sarkar et al., 2024). Faculty members are often selected based on disciplinary expertise, and as a result, may struggle to transform content knowledge into effective instruction without deliberate support systems (Brownell & Tanner, 2012; Gehrtz et al., 2022). As stated earlier, this challenge is particularly salient in data science, where there is limited consensus on both data science programs and the content and delivery of IDS courses.

To the best of our knowledge, aside from the argument of Hazzan and Mike (2023), there are no PCK studies y focused on IDS but PCK has been studied in data science related fields such as statistics, computer science, and domain-specific data science education. The following section provides a brief overview of PCK studies conducted in these fields.

*Pedagogical Content Knowledge in Data Science Education and Related Fields*

Several studies have examined how instructors approach the teaching of data-related content. Emery et al. (2021) found that biology and environmental science instructors valued skills such as data management, analysis, and visualization, while coding, modeling, and reproducibility were identified as areas where further training was needed. Similarly, Kross and Guo (2019) reported that practitioners who taught data science in academia or industry struggled with balancing abstraction and authenticity, selecting pedagogically appropriate datasets, and helping students from diverse backgrounds.

Other studies have focused on pedagogical tools and instructional reasoning. Sulmont et al. (2019) interviewed machine learning instructors teaching non-majors and identified elements of their PCK related to student misconceptions and teaching tactics, as well as difficulties in addressing more complex learning goals that involve integrating multiple concepts and applying skills to problem-solving.

In the context of computing and programming, Doukakis et al. (2010) documented gaps in teachers' pedagogical confidence despite relatively high content and technological knowledge. Complementing this, Saeli et al. (2011) reviewed literature on programming instruction and identified key components of PCK such as rationale, core concepts, student difficulties, and teaching strategies, often treated in isolation. This pointed to a need for more integrated research in informatics education. Similarly, research on introductory programming (Qian & Lehman, 2017) identified student difficulties related to syntactic, conceptual, and strategic knowledge and compiled teaching strategies and tools designed to address these misconceptions and other learning challenges. Extending these concerns to data science, Lau et al. (2022) observed that computer science and statistics instructors often embedded data science into existing courses but emphasized the complexity of designing dedicated courses that balance content from both domains while addressing diverse student backgrounds and limited instructional time.

In summary, efforts to explore the nature of PCK about data science and related disciplines are part of a growing literature focused on enhancing educators' ability to design courses and teach data-related content effectively. Our study contributes to these ongoing conversations by describing instructors' knowledge of IDS and by addressing gaps in what we have known about how instructors teach IDS courses.

## Methodology

### Research Design

In this study, we adopted qualitative research design and employed a *basic qualitative study* methodology (Merriam, 2009) to explore how IDS instructors describe and make sense of their teaching experiences. A basic qualitative study is used when the goal is to understand how individuals interpret their experiences, construct meaning, and describe their experiences within a particular context. This methodological approach is particularly well suited to examining teaching in IDS, a field that is still emerging and in which instructors often come from diverse disciplinary backgrounds. Additionally, definition of data science varies considerably across different institutions and departments. As a result, their interpretations of teaching practices and challenges warranted careful qualitative examination. By using this methodology, we aimed to develop a rich, contextualized understanding of instructors' perspectives, with Park and Oliver's (2008) PCK framework serving as an analytic lens for interpreting how instructors understand their teaching of IDS.

**Sample**

We identified our target population as instructors who taught an IDS course at least twice at the undergraduate level and whose course titles included Data Science and one of the keywords: *Introduction*, *Principles*, *Elements* or *Fundamentals*. We chose teaching an IDS course at least twice as a criterion because teaching a course for the first time often involves navigating basic instructional demands and developing an emerging teaching identity (Mutton et al., 2011; Pillen et al., 2013), whereas at least two iterations of the experience are likely to provide deeper reflections on both the course and their students. A total of 14 IDS instructors participated in the study.

To be able to provide a thick description of the context of IDS courses in our sample, we asked several background questions. These included: Students' major, class size, years of experience in teaching, whether having a teaching team or being the sole instructor, departments which offered the IDS courses, instructors' terminal degree and type of institution (research universities, liberal arts colleges) were recorded. Participant characteristics such as gender and age were not collected.

Participants were affiliated with different institutional types, including six research universities and eight liberal arts colleges. The instructors' experience teaching IDS courses ranged from 1 to 10 years ($M$ = 4.07, $Mdn$ = 3). Most instructors held doctoral degrees in mathematics or statistics-related fields, including mathematics, biostatistics, statistics, statistics and data analytics, and mathematics education. Two instructors held doctoral degrees in related disciplines (genetics and economics), and two held master's degrees in statistics or computer science.

The students enrolled in these IDS courses came from a variety of majors. While many IDS classes predominantly included students majoring in computer science, statistics, data science, or mathematics, they also attracted students from fields like cognitive science, engineering, social sciences, humanities, and undeclared majors[1]. Class sizes varied substantially across the courses, with nine courses enrolling fewer than 40 students and five courses enrolling more than 100 students. Instructional teams were more commonly observed in larger IDS courses, while smaller courses were often taught by a single instructor. IDS courses in the sample were offered across a range offering unit (Table 1).

[1] Some students in the US begin university without declaring a major and choose one within their first two years of study.

**Table 1**

*Academic Units Offering Introductory Data Science (IDS) Courses*

| Course Offering Unit | *n* |
|---|---|
| Mathematics | 3 |
| Mathematics and Statistics | 2 |
| Statistics | 2 |
| Center/Institute | 2 |
| Statistics and Data Science | 2 |
| Mathematics and Computer Science | 1 |
| Data Analytics | 1 |
| Other | 1 |

We also asked participants whether they had formal training in data science or pedagogy as well as their terminal degree to be able to understand the nature of the sample. These data are given in Table 2.

**Table 2**

*Formal Training in Data Science and Pedagogy**

| | Data Science | Pedagogy |
|---|---|---|
| Self-Taught | 4 | 0** |
| Workshops | 4 | 4 |
| Industry Experience | 2 | N/A** |
| Data Science Related Course Enrollment/Degree | 9 | N/A |
| Teaching Assistant Trainings, Course Enrollment, Internship | N/A | 8 |
| Terminal Degree | 0 | 1 |
| No Formal Training | 0 | 3 |

* *The column totals do not sum to 14 because some participants provided multiple answers.*

** *N/A indicates that the category is not applicable, whereas '0' indicates that no participants reported the given experience.*

As shown in Table 2, despite having no formal training in data science, the participants had training and terminal degrees from data science related disciplines such as computer science, statistics and mathematics, participating in workshops or self-taught. Most participants have no terminal degree in pedagogy; however, they reported participating in teaching workshops, teaching assistant trainings, internships, and a couple of graduate courses. Additionally, participants' terminal degrees were largely aligned with the academic units in which they taught (10 of 14).

**Instrumentation**

We adopted Park and Oliver's (2008) Hexagon model of PCK and developed instrument items tailored to the IDS context to examine IDS instructors' teaching experiences. The instrument consisted of 33 core interview questions encompassing the six PCK components (# of questions per component - orientations: 4; knowledge of student: 6; curriculum: 4; instructional strategies: 3; assessment: 5; teacher efficacy: 2; contextual: 9); follow-up questions were used as needed but are not included in this count. The semi-structured interview protocol is provided in Appendix A.

We established the content domain of the protocol via a three-step process. First, a comprehensive literature review was conducted on PCK research in (1) science education, (2) statistics education, and (3) data science education and related fields within higher education. This informed the scope of the interview questions and offered an initial understanding of potential IDS PCK components. Second, multiple collaborative meetings with the researchers of this study were held to refine the draft interview questions developed by the first author. These meetings focused on revising, adding, and tailoring questions aligned with the PCK framework. This iterative process ensured a broad range of questions aimed at capturing the nature of IDS instructors' PCK. Finally, the draft protocol was piloted with two IDS instructors. Feedback from the pilot study informed revisions, including adjustments to question phrasing and sequence, adding new questions for the main study, and length of the interview.

Both the pilot and main studies were conducted in compliance with ethical and institutional guidelines. The study received a Data Protection Registration Number and underwent a risk assessment in accordance with risk assessment procedures. This study was deemed exempt from ethics committee approval, as determined by the head of the department. All procedures involving human participants adhered to institutional ethical standards, and informed consent was obtained from all participants.

## Data Collection

We recruited the participants through mailing lists and online forums catering to large teacher-scholar communities serving statistics, computer science, and data science educator communities. Potential participants filled out a short survey where we asked (1) the name of the course that they teach, (2) whether that course is an undergraduate course or not; and (3) how many times they had taught the course. Based on these criteria, a pool of eligible instructors was identified, from which participants were purposefully selected using a criterion-based approach. The resulting sample reflected variation across institutional type, disciplinary background, and professional experience. Data were collected via online semi-structured interviews; while some selected instructors did not ultimately participate, thematic saturation was reached at 14 participants. Additional eligible instructors were available in the pool and would have been invited had new themes continued to emerge. Each interview lasted approximately 90 minutes, and participants received a gift card. At the time of the interviews, most participants were teaching in North America. Course syllabi were also collected, when available, to provide contextual information that supported interpretation of the interview data.

## Data Analysis

Data analysis was iterative throughout the study. To ensure participant anonymity, identifying information such as names, affiliations, and other unique details were removed during transcription. Consistent with *basic qualitative study* methodology (Merriam, 2009), we listed statements pertaining to the IDS teaching experience and employed *preliminary grouping* by using PCK elements guided by Park and Oliver's Hexagon Model of PCK for science teaching. While the overarching themes were informed by this framework, the analysis of the second element, Knowledge of Students' Understanding, was further guided by Qian and Lehman's (2017) classification of syntactic, conceptual, and strategic knowledge to examine student learning difficulties (see the Literature Review section for details).

A comprehensive codebook was developed to identify potential themes and categories, employing both inductive and deductive coding approaches. During the initial open coding phase, the three authors independently coded different interview transcripts to inductively identify patterns within each PCK element. The authors met to discuss emerging codes and compare interpretations, resulting in a shared codebook. After the codebook was established, the first author coded all fourteen interviews. To enhance analytic consistency, the second author independently coded two interviews, and discrepancies were discussed to ensure alignment in interpretation. The finalized codebook is provided in Appendix B.

During data analysis, course syllabi were examined to corroborate and contextualize instructors' descriptions of course content, structure, and instructional expectations. Syllabi were not coded independently; rather, they were used as supplementary instructional artifacts to support the interpretation of interview data. Specifically, we systematically examined syllabi through the lens of the PCK elements, attending to aspects such as sequencing of topics as indicators of curricular knowledge, assessment practices, and instructional orientations reflected in course design. In this way, the syllabi provided additional analytic insight into instructors' curricular decisions and elements of their PCK as reflected in planned instruction, complementing the interview data.

## Trustworthiness of the Study

To ensure trustworthiness in this qualitative study, we addressed credibility through *data triangulation*, *adequate engagement in data collection*, and *researcher reflexivity* (Merriam & Tisdell, 2015). Three researchers were involved in coding to support the dependability and confirmability of the findings. Transferability was addressed through the inclusion of *thick description* and *maximum variation*.

To support credibility, data were gathered from two sources, including semi-structured interviews and IDS course syllabi. While there is no universally accepted threshold for *adequate engagement in data collection,* achieving data saturation is a commonly suggested criterion. Data collection continued until no new information emerged. Fourteen participants contributed diverse perspectives on IDS teaching and learning environments.

As part of ensuring credibility through reflexivity, we disclose our areas of expertise and experience, which may have influenced the study's results. Our collective research interests span various aspects of data science education, possessing formal training in pedagogy, albeit in distinct areas. We also have experience in teacher education, particularly in preparing future STEM educators. One of us has been teaching an IDS course for many years. Notably, each of us brings unique, contextually shaped PCK (Carlson et al., 2019), which offers multiple perspectives on understanding PCK. Collectively, these shape our analysis of the nature of PCK among IDS instructors.

To enhance transferability, we provided thick descriptions of the IDS participants and their classroom environments. Furthermore, the participants in this study represent diverse backgrounds, including varying levels of teaching experience, industry involvement, institutional types, and terminal degrees. This diversity within our sample contributes to a rich portrayal of different IDS teaching contexts, enhancing the transferability of our findings.

## Findings

In line with conventions in PCK research in science, findings are presented as assertions, analytically generated claims grounded in participants' accounts (e.g., Friedrichsen et al., 2009; Park & Oliver, 2008). Based on the analysis of interview transcriptions and syllabi collected, eight major assertions were identified concerning the nature of the PCK of IDS instructors as follows:

- **Assertion 1:** IDS instructors' perceptions related to their purposes, goals, and reasons for teaching data science reflect three orientations. (Orientations to Teaching)
- **Assertion 2:** IDS students come to IDS courses with some preconceived notions which interact with their learning experience. (Knowledge of Students' Understanding)
- **Assertion 3:** IDS instructors observed varied student difficulties, including syntactic, conceptual, and strategic challenges. (Knowledge of Students' Understanding)
- **Assertion 4:** IDS instructors employ multiple instructional strategies, the majority of which can be categorized as subject-specific.(Knowledge of Instructional Strategies and Representations for Teaching)

- **Assertion 5:** Despite the absence of national guidelines, IDS instructors showed considerable overlap in early course content, though some felt less confident teaching more advanced/specialized data science topics. (Knowledge of Science Curriculum & Teacher Efficacy.)
- **Assertion 6:** The majority of IDS instructors used both formative and summative assessments, though only a few explicitly used these terms in describing their practices. (Knowledge of Assessment)
- **Assertion 7:** Conventional approaches to comparing and mapping PCK are not readily applicable in IDS contexts.
- **Assertion 8:** Background experiences inform IDS instructors' instructional decisions.

The first six assertions correspond to PCK components (Park & Oliver, 2008), whereas the final two emerged inductively from the interview data. Each of these assertions is examined in greater detail below.

**Assertion 1: IDS instructors' perceptions related to their purposes, goals, and reasons for teaching data science reflect three orientations.**

Participants articulated a variety of purposes, goals, and reasons for teaching IDS, which revealed three main instructional orientations: (1) enhancing data literacy/teach to learn from data; (2) familiarizing students with a programming language; and (3) attracting students to major/minor in data science. Among the 14 participants, nine referenced data literacy, two emphasized introducing students to a programming language, and four described the course as a strategic entry point into the data science major/minor. These orientations, however, were not mutually exclusive. Several instructors articulated goals that spanned multiple categories, reflecting the multifaceted nature of their teaching purposes.

This convergence of goals was articulated by one instructor who described how their teaching approach integrates several of these orientations:

> *Interviewee-09:* So, one of the main reasons I teach it is, as I mentioned, a little bit before to help students understand the field, right? It's still a new field. A lot of people talk about data science on our campus. I want students [in] this course to help them figure out if this is a possible path for them. I want them to come in and know that they don't already have to have taken [a] bajillion math courses. They don't have to be a programming wiz; they don't have to have been in [a] programming club in high school that they can start here…*[conversation continues]*... It is not meant to weed students out. It is meant to introduce the topics.

**Assertion 2: IDS students come to IDS courses with some preconceived notions which interact with their learning experience.**

Across all interviews, instructors reported that students often entered IDS courses with a range of preconceived notions about data science and programming. According to them, these ideas often interfered with how students engaged with the material and interpreted their own progress.

Instructors identified several stereotypes that interfered with student learning. Some students believed that proficiency in data science was about speed and accuracy in coding, leading to frustration when they encounter error messages. As one instructor explained, "*They are frustrated that it's not a simple task, very similar to math anxiety*." Others observed that students underestimated the iterative nature of programming: "*At first, they don't understand that programming is trial and error a lot.*"

Several instructors noted a fixed mindset in some students, who assumed they were either naturally good or bad at data science. Instructors suggested that these misunderstandings may stem from external influences, including cultural narratives, conveyed through such things as media portrayals, YouTube tutorials, and societal expectations.

**Assertion 3: IDS instructors observed varied student difficulties, including syntactic, conceptual, and strategic challenges.**

Ten participants described difficulties related to students' syntactic knowledge, such as understanding programming structure and syntax. A common observation among instructors was that students without prior programming experience required significantly more time to become comfortable with coding. Thirteen instructors noted conceptual challenges, particularly in mathematical or statistical reasoning. In addition, eleven instructors emphasized strategic difficulties related to students' ability to integrate programming skills with disciplinary understanding to solve novel problems in data science projects.

**Assertion 4: IDS instructors employ multiple instructional strategies, the majority of which can be categorized as subject-specific.**

All participants in this study reported employing more than one instructional strategy in their IDS classrooms. Upon classification, subject-specific strategies were more frequently used than were topic-specific ones, particularly in larger IDS courses, whereas smaller courses tended to benefit from topic-specific approaches. The most commonly reported strategies are presented in Table 3.

**Table 3**

*Most Commonly Used Subject-Specific Instructional Strategies*

| Subject Specific Teaching Strategies | Number of Instructors (*n*) |
|---|---|
| Lecturing | 9 |
| Live Coding | 8 |
| Questioning | 3 |
| Group Work | 3 |
| Think-Pair-Share | 3 |

Some instructors described using instructional approaches tailored to specific sub-topics. For instance, storytelling was employed when introducing real-world cases; questioning was used to facilitate discussions around data ethics; role playing was incorporated in teaching data joining concepts; and tactile simulations were used to illustrate sampling distributions. The use of role playing was elaborated upon by one instructor as follows:

> *Interviewee-07:* In the unit where we talk about joining two tables together, actually, my colleague came up with us who also has a background in theater. And so [he/she/they] came up with this pretty cool thing that I've been using ever since, where you get students to come up at the beginning of the class. And you have… We have these little sorts of placards that they get to wear that have data values on them. And you sort of line them up like two tables. And then you illustrate sort of what happens when there's a left join and, you know, the students have to like come together, and like they have a partner, or maybe they don't maybe there's a missing value, or maybe there's not and so you can illustrate the difference between like a left join, a right join, and inner join like physically, with students in the front of the room role playing or something.

This example showcases how some instructors incorporate topic-specific teaching methods to support students' understanding of abstract concepts. While less common than subject-specific strategies such as lecturing and live coding, these approaches were typically tied to the instructional needs of a particular concept.

**Assertion 5: Despite the absence of national guidelines, IDS instructors showed considerable overlap in early course content, though some felt less confident teaching more advanced/specialized data science topics.**

To examine content consistency across IDS classrooms, we analyzed course syllabi collected from the participants. In the absence of a centralized curriculum framework or a guiding document, variation in content selection was expected. These are summarized in Table 4.

**Table 4**

*Topics Taught in IDS*

| Topic | Number of IDS instructors |
| --- | --- |
| Introduction to Data Science | 14 |
| Data and Variables | 14 |
| Data Visualization | 14 |
| Data Wrangling | 14 |
| Statistical Inference | 7 |
| Ethics | 6 |
| Introduction to Machine Learning | 4 |
| Text Analysis | 4 |
| Clustering Analysis | 3 |

As shown in Table 4, IDS instructors demonstrated considerable overlap in early course content. This overlap was observed across academic units; however, for more advanced or specialized topics, no clear pattern emerged between academic units and topics taught.

Instructors also made varied decisions regarding the programming tools used for instruction. Seven instructors reported using R as the primary programming language, while two used Python. Two instructors integrated both R and Python, and two others used a combination of SQL (Structured Query Language) and R. One instructor reported not incorporating any programming language in their instruction.

Self-efficacy regarding IDS topics and pedagogy was another PCK element we explored. Thirteen participants expressed high self-efficacy in foundational IDS topics, such as data wrangling, data visualization, and debugging. However, seven participants reported feeling less confident when teaching content, especially advanced or specialized data science topics like machine learning, big data management, and text analysis.

Taken together, these findings suggest that while no formal curriculum framework guides IDS course design, a shared foundation has organically emerged across instructors, particularly in the early stages of the curriculum. Variation increases in the second half of the course, particularly in relation to advanced topics and tool selection.

**Assertion 6: The majority of IDS instructors used both formative and summative assessments, though only a few explicitly used these terms in describing their practices.**

Although summative assessment was more commonly discussed, elements of both formative and summative assessment were present in most participants' instructional approaches. A total of 13 instructors reported using summative assessments, and six described practices consistent with formative assessment. One instructor mentioned using an ungrading approach (Blum, 2020). However, only a small number of instructors explicitly referred to their assessments as “formative” or “summative.”

Summative assessment practices, as identified through both interview data and course syllabi, included homework assignments, quizzes, reading assignments, peer evaluations, exams, and final projects. Formative assessment practices, while less explicitly labeled, were evident in several instructors’ classroom routines. These included selecting individual students to participate in live coding with the instructor, asking students questions during instruction, and providing detailed, individualized feedback outside of class time. In terms of providing feedback to students, larger IDS courses tended to rely more on automated forms of feedback and instructional teams, whereas smaller courses benefited from instructors' office hours and oral feedback. During interviews, only one participant explicitly mentioned using Bloom's Taxonomy when assessing student learning.

**Assertion 7: Conventional approaches to comparing and mapping PCK are not readily applicable in IDS contexts.**

The analysis revealed a variation in academic units (Table 1), and their academic and professional backgrounds (Table 2), as well as instructional decisions complicating efforts to delineate and compare their PCK using conventional comparative frameworks. As summarized in Table 2, while nine instructors held graduate degrees in data science–related fields, none had received their degrees from dedicated data science departments or programs. Furthermore, only one participant had a terminal degree in pedagogy. The remaining participants had engaged with pedagogy through alternative means such as teaching assistant (TA) training, teaching internships, or coursework. In other words, variability in instructors' content and pedagogical knowledge bases translated into distinct configurations of PCK, limiting the utility of straightforward comparative approaches examining PCK across IDS instructors.

Additionally, most instructors had developed their teaching practices through workshops, industry experience, or self-directed learning. These varied pathways contributed to the development of highly individualized PCK profiles, with each instructor drawing from a unique mix of disciplinary expertise, pedagogical exposure, and teaching experience. This diversity was further amplified by contextual factors including the heterogeneity of IDS student populations, variations in class sizes, and institutional calendar structures (e.g., semesters vs. quarters). These factors appeared to play a role in instructional decision-making and the design of IDS curricula, adding another layer of complexity to understanding and comparing instructors' PCK. Taken together, these variations limited the possibility of making direct or standardized comparisons across instructors' PCK profiles in IDS settings.

**Assertion 8: Background experiences inform IDS instructors' instructional decisions.**

Participants brought a range of experiences to their roles as IDS instructors. These experiences — whether self-taught, industry-based, community-engaged, or shaped by departmental contexts— played a critical role in how they designed and delivered their courses.

For example, three participants had industry experience prior to becoming faculty members. They reflected on the disconnect between industry demands and traditional undergraduate coursework, which motivated them to align their courses more closely with real-world applications. When asked how their experiences in data science shaped their teaching, one instructor responded:

> *Interviewee-07:* Tremendously. I mean in fact; it's one of the main reasons that we have the course in the first place. … It was very clear to me sort of right away, that sort of what we were teaching really didn't match up with what students needed to know, and it's not that what we were teaching wasn't valuable, it was! … So, you know they were using modern tools, but they couldn't like… join 2 data sets together… But yeah, it was very much, you know, me having had that experience working in industry knowing the kinds of tasks that you have to do as a professional data scientist and seeing the gaps with what we were teaching and that made up much of the content of the data science course.

Some participants took the initiative to build their own competence. For example, four instructors reported using their sabbaticals or personal time to teach themselves IDS concepts. One such instructor explained how this learning journey shaped their pedagogical stance as *"…in some sense, I think that makes me a better instructor because it helps me relate to students a little bit… "*

Community engagement, such as participating in national conferences, workshops or peer discussions, was mentioned by several instructors in relation to their teaching practices. A total of six instructors described collaborating with community or other faculties in their universities. For instance, one participant stated that "*… before I taught data science, you know, I went to a spectrum of conferences*

*[giving some conference names]. I saw a lot of talks about some tools and… and resources people are using to teach data science, and I end up using, I ended up using a few of those resources…* "

These findings highlight the role of experiential learning and external engagement in shaping the PCK of IDS instructors, particularly in the absence of formal preparation pathways.

# Discussion and Conclusion

This study examined the nature of PCK among instructors teaching undergraduate IDS courses. To better understand this phenomenon, situated within the absence of formal preparation in content, pedagogy, and curricular guidelines, we employed a basic qualitative study. As a qualitative study, our aim is not to generalize but to offer in-depth insights into how instructors describe and make sense of their experiences teaching IDS. With this context in mind, we now turn to a discussion of our findings and what they reveal about how PCK can be characterized within IDS, a context in which it does not readily lend itself to simple comparison or standardization.

**Discussion on the Nature of PCK about IDS Teaching**

Perhaps the most important question is whether our study revealed elements of PCK drawn from the framework (Park & Oliver, 2008) we used to shape our interview protocol that could serve as a starting point for developing a PCK framework specific to IDS. While the question is not easily reduced to a binary response, we believe our findings offer empirical insights into the nature of PCK as it manifests in the context of teaching IDS. Before diving into that discussion, it is important to note that we intentionally restrict our discussion to PCK about IDS rather than proposing a generalized PCK framework for data science as a whole, given the specific scope of our research and the lack of a widely agreed-upon definition of data science as an emerging field. It is important that readers do not overextend these insights into the broader field of data science education.

*Elements of PCK about IDS Teaching*

It was not surprising that the six elements of Park and Oliver's (2008) Hexagon model of PCK appeared in our data, as we used them as the main themes of the codebook. However, the categories we identified within each theme revealed both convergence and divergence: some categories closely matched the original definitions, while others reflected dimensions unique to the IDS context.

PCK elements such as teacher self-efficacy, knowledge of curriculum, instructional strategies, and assessment were clearly evident in instructors' accounts. Prior research across international contexts in computing, statistics, and machine learning has often examined these dimensions individually. In terms of self-efficacy, Emery et al. (2021) reported that instructors emphasized foundational data skills while expressing lower confidence in more advanced areas. Similarly, Doukakis et al. (2010) documented gaps in pedagogical confidence despite relatively strong content knowledge. Our findings reflect comparable patterns. IDS instructors incorporated central data concepts into their syllabi while acknowledging varying levels of confidence in teaching complex topics. This suggests that tensions between content coverage and instructor self-efficacy may recur across data-related disciplines.

The absence of formal curricular guidelines was reflected in convergence in the early part of the course and divergence in the later across instructors. This raises important questions: Are these differences in content emphasis a result of instructors' disciplinary backgrounds (e.g., a computer science instructor including more programming-focused topics), or do they reflect differing interpretations of what an introductory data science course should be? While this was a plausible consideration, we did not observe a consistent pattern between instructors' academic/professional backgrounds, academic units and the nature of their curricular knowledge. As such, these questions remain open for future investigation.

At the same time, certain features of the IDS context became visible in knowledge of instructional strategies. Kross and Guo (2019) described challenges in balancing abstraction and authenticity in data science instruction. In our study, instructors with industry backgrounds described navigating this balance by articulating topic-specific pedagogical strategies for technically demanding concepts. In this sense, our findings align with prior scholarship identifying key some of the PCK components across computing-related fields (Saeli et al., 2011), while situating them within the particular framework of IDS.

The remaining two elements of PCK —orientations toward teaching IDS and knowledge of students' understanding—revealed distinct patterns that may reflect the particular demands and characteristics of the IDS teaching context. With respect to orientations toward teaching IDS, we identified three common themes. While goals such as promoting literacy and recruiting students may align with orientations found in other introductory courses in higher education, the focus on programming fluency stands out as particularly relevant to data science and/or its adjacent fields, such as statistics and computer science. This emphasis marks a departure from science education PCK, where science teachers' orientations often center on inquiry-based learning, conceptual understanding, and scientific reasoning (Friedrichsen et al., 2011). These distinctions suggest that orientations toward teaching IDS may require a different interpretive lens than those traditionally applied in science education contexts.

Knowledge of students emerged as a richly coded element of PCK in our study (Assertions 2, 3). Instructors frequently noted the wide range of student backgrounds in IDS classrooms, which posed significant challenges for instructional pacing, scaffolding, and other instructional decisions. Student learning difficulties were another dominant category. While we drew on a framework from computer science education (Qian & Lehman, 2017) to analyze patterns—such as syntactic, conceptual, and strategic knowledge difficulties—we observed characteristics that appeared specific to the interdisciplinary nature of IDS courses. We discussed these patterns in more detail in a separate conference paper (Demirci et al., 2023). Overall, these were consistent with previous studies on machine learning and programming education where these studies reported similar recurring student difficulties in addressing more complex learning goals (Sulmont et al., 2019; Qian & Lehman, 2017).

In addition, several instructors noted that students enter the course with stereotypical views of what it means to be a data scientist, often influenced by media and society. This pattern parallels findings from the “draw-a-scientist” literature in science education (Chambers, 1983), in which student perceptions of disciplinary identity interfere with their understanding of the subject and their place within it.

Together, these observations pertaining to the PCK of IDS instructors underscore its multifaceted and context-dependent nature, supporting our seventh and eighth assertions. Across our data, all six elements of PCK varied across instructors, reflecting the diversity of backgrounds, teaching contexts, and institutional settings. This variation contributed to the complex teaching experiences we recorded: experiences that resisted direct or standardized comparisons across instructors’ PCK profiles in IDS settings.

Yet, this lack of uniformity does not indicate the absence of structure or insight. Instead, it illustrates how PCK is actively constructed and enacted in response to the evolving, interdisciplinary nature of data science and in the absence of standardized training or frameworks. We observed both individualized PCK shaped by personal trajectories and emerging collective PCK fostered through shared teaching environments, consistent with prior characterizations of PCK as personal, dynamic, and idiosyncratic (Grossman, 1990; Van Driel et al., 1998; Park & Chan, 2025). Taken together, these findings suggest that even in the absence of explicit awareness or formal articulation of PCK, instructors are actively developing PCK about IDS through local practices, community exchange, instructors’ background and the demands of a rapidly expanding field.

## Recommendations and Implications for Practice

Our findings point to several opportunities to support the development of PCK for IDS instruction.

***Recommendation 1 - Develop a Shared Pedagogical Language:***

Our findings reveal that while instructors use a variety of effective teaching and assessment strategies, they often do not use formal pedagogical terms to describe them. For instance, Assertion 6 shows that the majority of instructors employed practices consistent with formative and summative assessment but only a few used these specific terms. A shared pedagogical language refers to the explicit and consistent use of established terminology by training instructors to identify and name practices such as “formative assessment,” “scaffolding,” and “think-pair-share,” thereby supporting clearer communication among educators. It would also bridge the gap between their intuitive, experiential knowledge and established educational research, enhancing collaborative reflection and professional development.

***Recommendation 2 - Target Professional Development to Specific PCK Needs:*** Our study found that instructors develop their pedagogical content knowledge through highly informal and individual pathways (Assertions 7, 8) and that they face specific, recurring challenges, such as managing diverse student backgrounds (Assertion 2) and teaching advanced topics (Assertion 5) where instructor confidence is lower. Therefore, we recommend that structured training and professional development be targeted to these specific needs.

***Recommendation 3 - Formalize the Emerging Core Curriculum:***

Assertion 5 demonstrates that, despite the absence of national guidelines, an informal consensus on core topics has organically emerged in IDS courses. This "emerging core curriculum," as identified in our syllabi analysis (Table 4), consistently includes an introduction to data science, data and variables, data wrangling, and data visualization. We recommend that professional organizations formally recognize and codify this common core. Doing so would provide a stable curricular foundation for new instructors and allow for the development of targeted, modular resources for the more variable and advanced topics that instructors felt less confident teaching.

**Conceptual and Methodological Reflections for Future Research**

In this study, Park and Oliver’s (2008) Hexagon model was used as an analytical framework to examine IDS instructors’ PCK. Thus, the identification of six PCK elements in our findings is therefore closely connected to this framework-guided approach. While the Hexagon model offered a coherent structure for organizing IDS instructors’ PCK, alternative conceptualizations of PCK may foreground different dimensions of instructional practice. Future research may employ alternative PCK frameworks to revisit the dimensional structure of PCK in IDS contexts, potentially deepening, expanding, or refining the elements identified here.

At the same time, we did not attempt to systematically map participants' PCK components or compare PCK across instructors, not only because of the breadth of the domain, but also due to substantial variation in disciplinary backgrounds, standardized knowledge, pedagogical preparation, and institutional contexts. Future research seeking to map relationships or interactions among PCK elements may benefit from adopting a topic-specific PCK lens, as a more bounded content focus could allow for more systematic analysis than was possible in the present study.

*Content Knowledge, Pedagogical Knowledge, and Collective PCK in IDS: Beyond the Scope, Within the Conversation*

Although not the primary focus of this study, patterns identified invite further consideration of content knowledge, pedagogical knowledge, and collective PCK in IDS. Variation in instructors' backgrounds is reflected in differences in instructional decisions. More broadly, variation appears across institutional locations of IDS programs (e.g., terminal degrees housed in different departments), student backgrounds, and the absence of consensus regarding what constitutes data science as a field. Together, these conditions complicate efforts to systematically conceptualize content knowledge in IDS, which in turn makes it difficult to compare PCK across instructors in a coherent manner, given the close relationship between content knowledge and PCK ( Shulman, 1987; Carlson et al., 2019). These patterns suggest the need for further research on how content knowledge in IDS can be systematically conceptualized for research and instructional purposes.

Pedagogical knowledge was not an explicit focus of this study; however, instructors' accounts reflected experience-based understandings of teaching and assessment that may align with established pedagogical practices. While some participants explicitly used pedagogical terminology (e.g., formative assessment), others described practices that closely without naming them as such. This distinction suggests that differences may lie less in the presence of pedagogical knowledge and more in how that knowledge is articulated. We also acknowledge that teaching remains a complex professional practice that extends beyond experience alone (Jackson, 1986). Further studies should examine how experiential knowledge and formal pedagogical preparation interact in shaping and articulating PCK in IDS contexts.

Finally, shared practices emerging through professional communities point to the potential development of collective PCK. Further studies should explore how such collective knowledge forms and stabilizes over time within IDS and data science education broadly.

### Limitations

One limitation concerns the lack of consensus about what constitutes data science or an IDS course. In response to this ambiguity, we used three pragmatic criteria to focus on courses commonly framed as introductory. Recruitment occurred through mailing lists and online communities centered on teaching, where instructors fitting these criteria were most likely to be found. As a result, participants may have had a particular interest in PCK for IDS instruction and may have been teaching courses more consistent with our selection criteria. While this approach enabled us to access in-depth perspectives, it may also have resulted in a sample of instructors who were more confident or stable in their instructional roles and whose courses aligned more closely with our selection criteria, potentially excluding instructors teaching similar content under different course titles or conceptualizations of data science.

Another limitation concerns the geographic scope. Although we did not restrict recruitment geographically, all participants were based in North America. This may reflect the influence of our selection criteria and outreach channels, which could have unintentionally limited participation from instructors in other educational systems. Thus, the findings should be understood within the cultural and institutional context of North American higher education. Instructors teaching similar courses in other countries may face different structural, curricular, or cultural challenges. Future research in diverse educational settings could provide additional insights and deepen our understanding of how PCK about IDS manifests across contexts.

Lastly, because we adopted Park and Oliver's (2008) PCK framework from science education, our analysis may not have captured dimensions of PCK that are unique to IDS. As a result, certain aspects of IDS teaching may have remained underexamined within the boundaries of this model. PCK frameworks developed in other disciplines as well as technology-integrated PCK frameworks could also offer additional perspectives on IDS PCK; however, including these frameworks was beyond the scope of this single study.

## Declaration

### Availability of data and materials

The datasets generated and/or analyzed during the current study are not publicly available in order to protect participant confidentiality, as the data were collected through interviews and cannot be fully de-identified. However, the codebook and selected illustrative deidentified statements are provided in the Appendix and within the manuscript.

### Human Ethics and Consent to Participate

This study was deemed exempt from ethics committee approval, as determined by the head of the department at UCL. Both the pilot and main studies were conducted in compliance with ethical and institutional guidelines. The study received a Data Protection Registration Number and underwent a risk assessment in accordance with risk assessment procedures. All procedures involving human participants adhered to institutional ethical standards, and informed consent was obtained from all participants.

### Authors' contributions

All authors conceptualized the study. SD and MD developed the methodology, verified the reproducibility of the results and secured financial support for the project leading to this publication. SD, MD, and JMR curated data. SD coordinated and administered the project, conducted the research, collected and analyzed the data. All authors contributed to writing the original draft and to reviewing and editing the manuscript.

**Funding**

Demirci has been supported by TUBITAK 2219 - International Postdoctoral Research Fellowship Program for Turkish Citizens. Demirci and Doğucu completed an earlier part of this work in the Department of Statistical Science at University College London and have been supported by University College London, Mathematical and Physical Sciences. Doğucu has been supported by NSF IIS award #212336. Rosenberg has been supported by NSF Grant No. 2239152.

**Competing Interests**

The authors declare no competing interests.

**Acknowledgements**

Not applicable

**Declaration of generative AI and AI-assisted technologies in the writing process**

During the preparation of this work, the first author occasionally used ChatGPT to improve language and readability. After using this tool, the authors reviewed and edited the content as needed and take full responsibility for the content of the publication.

## APPENDIX A

### SCRIPT PRIOR TO INTERVIEW:

***[small talk for ice-breaking]***

I'd like to thank you once again for being willing to participate in our study. As I have mentioned to you before, our study seeks to understand the Pedagogical Content Knowledge (PCK) of the instructors giving the course of introductory data science.

Before starting, let me introduce myself briefly. *[The interviewer introduces themselves.]*

*(Optional if requested)* PCK could be briefly defined as following: "*PCK is teachers' understanding and enactment of how to help a group of students understand specific subject matter using multiple instructional strategies, representations, and assessments while working within the contextual, cultural, and social limitations in the learning environment*." (Park & Oliver, 2008, p.264)

*(Optional if requested)* The aim of this research is to develop a measurement tool that measures PCK of the Introductory Data Science (IDS) instructors and this part is considered as a qualitative part of the research to be able to collect data to develop this tool.

I would like to remind that there is no wrong answer, we just want to hear your experiences, thoughts, and comments on IDS teaching. Our interview today will last approximately one and a half hours during which I will be asking you about your conception about data science; a typical week in your IDS course and elaborating it in terms of orientations towards IDS teaching, the curriculum for data science, the students' characteristics, your assessment strategies, and your instructional strategies.

***[review aspects of consent form, confirm that they completed the form]***

Previously, you completed a consent form indicating that I have your permission (or not) to video and audio record our conversation.

Are you still ok with me recording (or not) our conversation today? ___Yes ___No

**If yes:** Thank you! **I will now turn on the recording.** Please let me know if at any point you want me to turn off the recorder or keep something you said off the record.

**If no:** Thank you for letting me know. We will finish our meeting now. Thank you for your time.

***[turn-on recording]***

Before we begin the interview, do you have any questions? ***[Discuss questions]***

If any questions (or other questions) arise at any point in this study, you can feel free to ask them at any time. I would be more than happy to answer your questions.

If you want to have a break, you can have it anytime you want. You can simply ask for it.

| | ***Upbringing:*** To begin this interview, I'd like to ask you some general questions about Data Science and your experiences with Data Science |
|---|---|
| | **Questions** |
| ☐ | **1.** What does «data science» mean for you, in general?<br>**<u>Follow-up:</u>**<br>i. Which field(s) are part of data science, in your opinion?<br>ii. What topic(s) of these fields are part of data science, in your opinion? |
| ☐ | **2.** Could you describe your experiences with data science?<br>**<u>Follow-up:</u>**<br>i. How have these experiences shaped your teaching of DS?<br>ii. How has your teaching of DS changed over time? |
| ☐ | **3.** Could you please share your opinion about the purposes and goals for teaching data science?<br>**<u>Follow-up:</u>**<br>i. Why do we teach data science?<br>ii. How do these goals and purposes inform your teaching in data science? |
| | ***Contextual Information of the IDS Course:*** Thank you for your responses. Now, I'd like to ask you some general questions about the context of your IDS course. |
| ☐ | **4.** How long have you been teaching in this course? |
| ☐ | **5.** What is your primary role in the IDS course?<br>i. How many staff are responsible for this course? Are there any TA / graders /co-instructors in your course? |
| ☐ | **6.** From which departments and majors do students enroll in your IDS course? |
| ☐ | **7.** How many students (approximately) are enrolled in your IDS course? |
| ☐ | **8.** Does the IDS course you teach have prerequisites? Is the IDS course that you teach is a prerequisite to any other courses?<br>**<u>Follow-up:</u>**<br>i. How does this impact what you include or exclude in your course?<br>ii. How does this impact sequencing of topics within your IDS course? |
| ☐ | **9.** How interdisciplinary are the topics in your IDS course?<br>**<u>Follow-up:</u>** How connected to other disciplines is the teaching of your IDS course? |
| ☐ | **10.** Could you please share your opinion about the purposes and goals for teaching IDS? |
| | **Follow-up:**<br>i. Why do we teach IDS?<br>ii. How do these goals and purposes inform your teaching in IDS? |
| | Thank you for sharing this general information about the profile of your IDS course. Now, I’d like to hear about the flow of your typical week in your IDS course.<br><br>**Note to the interviewer:** *Question 10 and beyond are elaborative questions to ask in case of not having comprehensive answers for each component of PCK. You can skip the questions which the interviewee has already answered.* |
| ☐ | **11.** What does your typical week in your IDS course look like?<br>**Follow-up:**<br>i. What are students doing in class?<br>ii. What do they do outside of class?<br>iii. How do you prepare (ideally)? |
| ☐ | **12.** How do you include technology to your IDS course? |
| | **ELABORATIVE QUESTIONS** |
| | ***IDS Curriculum:*** Thank you for your responses. In the previous questions, we have already started to talk about the IDS curriculum. Now, I’d like to ask you a few elaborative questions regarding the IDS curriculum of your course. |
| ☐ | **13.** What--if anything--is important for students to already know when they begin your course? |
| ☐ | **14.** Do you develop the content and materials for your course?<br>**IF YES:**<br>i. How did you go about developing the content or materials?<br>a. What resources-if any-did you draw on while developing the content and materials for your course?<br>b. Do you rely on any curricular guidelines while planning your IDS course?<br>**IF YES:** How?<br>**IF NO**: How would you modify any content/material? |
| | ***IDS Learners:*** Thank you for sharing information about the IDS curriculum. I’d like to now ask you some elaborative questions about the students enrolled in your IDS course. |
| ☐ | **15.** Could you please share if and how you adapt your teaching to different learners from different departments and/or majors? |
| ☐ | **16.** With which concept(s) do IDS students have difficulties? |
| ☐ | **17.** What are the difficulties, if any, that students have while performing DS tasks given in your IDS course? (*We could give an example here*) |
| ☐ | ***18.*** What are the misconceptions or stereotypes about DS or data scientists that you have observed in IDS students? (*misconceptions/stereotypes about learning or doing DS)*<br><br>**<u>Follow-up:</u>**<br><br>i. What could be the sources for these conceptions?<br>ii. How do you detect and address them?<br>iii. Do they prevent them from learning? **<u>IF YES:</u>** How? |
| | **IDS Instructional Strategies:** Thank you. I'd like to now ask you a few questions specifically about your IDS instructional strategies |
| ☐ | **19.** What are the teaching methods or strategies you use to teach IDS?<br><br>**<u>Follow-up:</u>**<br><br>i. What are the teaching methods or strategies you use to teach conceptual parts of your IDS course? (*e.g., concept of sampling distribution, how to interpret data or drawing conclusions*)<br>ii. What are the teaching methods or strategies you use to teach technical parts of your IDS course? (*e.g., how to use software programs, how to write code*)<br>iii. How--if at all--does technology relate to the teaching strategies you use? |
| ☐ | **20.** Do you use any specific teaching strategies when you notice that students have difficulties with particular ideas or topics? |
| ☐ | **21.** What kind of teaching approaches – if any – would you use to avoid creating misconceptions? |
| | **IDS Assessment:** This set of questions are focused on getting to know more about your IDS Assessment strategies |
| ☐ | **22.** How do you assess learning in your introductory data science classroom/lab session/recitation? |
| ☐ | **23.** Which assessment strategies do you use in your IDS class?<br><br>**<u>Follow-up</u>:** What purpose do they serve? |
| ☐ | **24.** When and how often do you assess your students?<br><br>**<u>Follow-up</u>:** For what reason do you assess your students at these times/in these ways? |
| ☐ | **25.** What kind of feedback –if any-- do you provide your students?<br><br>**<u>IF YES</u>:** How often? |
| ☐ | **26.** What is your overall grading policy?<br><br>i. Could you please share, if any, how you grade an assignment/exam in your IDS course? |

| | ii. How much does each assessment strategy contribute to a student’s overall grade? |
|---|---|
| | My final set of questions is about your self-efficacy beliefs towards IDS and IDS teaching. Let me remind you that you are free not to respond to any question that you do not feel comfortable answering. |
| ☐ | **27.** How confident do you feel about your understanding and knowledge related to IDS concepts? Are there concepts/topics you feel more or less comfortable with? |
| ☐ | **28.** Are there certain pedagogical aspects of teaching IDS you feel more or less comfortable with? |
| | Before we conclude this interview, is there something about your experience in this IDS teaching that you think it is worth sharing that we have not yet had a chance to discuss? |
| | **Additional Information of the Participant:** Thanks for your contribution. Last, if you agree to answer, I’d like to ask you about your position/experiences. Let me remind you that you are free not to respond to any question that you do not feel comfortable answering. |
| ☐ | a) Which department are you currently affiliated with? |
| ☐ | b) Do you have any formal training in data science? *(If the interviewee asks for a definition for data science, the interviewer will leave it to the interviewee)* |
| ☐ | c) Do you have any formal training in teaching? |
| ☐ | d) What is your terminal degree in what discipline? |

| **APPENDIX B: CODEBOOK** | | |
| --- | --- | --- |
| **Themes** | **Categories** | **Sub-categories + Codes** |
| **Theme 1:**<br>Data Science (DS) | **Category 1.1:** Definition of DS | ***Sub-category 1.1.1: Data Oriented Definitions***<br>**DS111-01:** data-based decision making<br>**DS111-02:** understanding data<br>**DS111-03:** answering interesting questions with data<br>**DS111-04:** A set of practices and algorithms to process data and get insight from data |
| | | ***Sub-category 1.1.2: Discipline Oriented Definitions***<br>**DS112-01:** marriage between CS and statistics<br>**DS112-02:** a combination/intersection of math, statistics, and CS<br>**DS112-03:** intersection of the data side of statistics, computer science and some domain knowledge<br>**DS112-04:** a combination of math, statistics, CS, and other domains<br>**DS112-05:** a combination of statistics and programming |
| | | ***Sub-category 1.1.3: Definition in Progress***<br>**DS113-01:** still being defined |
| | **Category 1.2:** Fields in DS | **DS12-01:** statistics<br>**DS12-02:** computer science<br>**DS12-03:** mathematics<br>**DS12-04:** subject matter /domain knowledge<br>**DS12-05:** biological sciences<br>**DS12-06:** social sciences<br>**DS12-07:** economics<br>**DS12-08:** psychology<br>**DS12-09:** business analytic<br>**DS12-10:** difficult to establish boundaries<br>**DS12-11:** political science<br>**DS12-12:** humanities |
| | **Category 1.3:** Topics of DS | ***Sub-Category 1.3.1: Statistics_Topics***<br>**DS131-01:** databases<br>**DS131-02:** data visualization<br>**DS131-03:** machine learning |
| **Theme 1:**<br>Data Science (DS)<br>***(Cont'd)*** | **Category 1.3:** Topics of DS ***(Cont'd)*** | **DS131-04:** deep learning<br>**DS131-05:** statistical thinking<br>**DS131-06:** confounding variables<br>**DS131-07:** causal inference<br>**DS131-08:** regression techniques<br>**DS131-09:** model validation<br>**DS131-10:** statistical modelling<br>**DS131-11:** CART and random forest<br>**DS131-12:** PCA<br>**DS131-13:** sampling design<br>**DS131-14:** modeling<br>**DS131-15:** processing data<br>**DS131-16:** multivariate statistics<br>**DS131-17:** inferential statistics<br>**DS131-18:** clustering<br>**DS131-19:** data wrangling<br>**DS131-20:** everything in statistics<br>**DS131-21:** understanding of human behavior<br>**DS131-22:** simulations |
| | | ***Sub-Category 1.3.2: Computer Science_Topics***<br>**DS132-01:** computational thinking<br>**DS132-02:** algorithmic thinking<br>**DS132-03:** coding<br>**DS132-04:** programming<br>**DS132-05:** database structures<br>**DS132-06:** data visualization<br>**DS132-07:** machine learning<br>**DS132-08:** deep learning<br>**DS132-09:** information management<br>**DS132-10:** database storage<br>**DS132-11:** data engineering<br>**DS132-12:** data mining |

| **Theme 1:**<br>Data Science (DS)<br>***(Cont'd)*** | **Category 1.3:** Topics of DS ***(Cont'd)*** | |
|---|---|---|
| | | ***Sub-Category 1.3.3: Mathematics_Topics***<br>**DS133-01:** problem solving<br>**DS133-02:** analytical thinking<br>**DS133-03:** probability theory<br>**DS133-04:** linear algebra<br>**DS133-05:** information theory<br>**DS133-06:** statistics (a branch of math)<br>**DS133-07:** calculus<br>**DS133-08:** optimization<br>**DS133-09:** dynamical systems<br>**DS133-10:** stochastic processes<br>**DS133-11:** algorithms |
| | | ***Sub-Category 1.3.4: Subject Matter Expertise_Topics***<br>**DS134-01:** each field's topics that related to data science<br>**DS134-02:** Communication<br>**DS134-03:** Data Ethics |
| | **Category 1.4:** Purposes, goals, and reasons for teaching DS | ***Sub-Category 1.4.1: Student Growth Oriented Reasons***<br>**DS141-01:** to make students aware of the power of decision making or knowledge gaining from data.<br>**DS141-02:** to enable students to critically evaluate scientific literature from the perspective of data analysis<br>**DS141-03:** to enhance the data skills of students in other fields<br>**DS141-04:** to enable students to write and communicate effectively as a data scientist.<br>**DS141-05:** to convince students to choose data science as a major.<br>**DS141-06:** to develop a sense of agency<br>**DS141-07:** to give a set of broadly transferable quantitative skills<br>**DS141-08:** to enable students to become good decision making,<br>**DS141-09:** to enable students to become a better citizen<br>**DS141-10:** to teach them to learn things from data<br>**DS141-11:** every person in 20 more years needs to be able to actively work with the data set in some way. |

| **Theme 1:**<br>Data Science (DS)<br>***(Cont'd)*** | **Category 1.4:** Purposes, goals, and reasons for teaching DS ***(Cont'd)*** | **DS141-12:** to teach how to do data science responsibly. |
|---|---|---|
| | | ***Sub-Category 1.4.2: Market Demand Oriented Reasons***<br>**DS142-01:** to train future data scientists<br>**DS142-02:** to solve problems with data<br>**DS142-03:** It is very popular.<br>**DS142-04**: data are everywhere. |
| | **Category 1.5:** Experience in DS | ***Sub-Category 1.5.1: Community Engagement Experience***<br>**DS151-01:** DS Curriculum Design Team Member<br>**DS151-02:** Participation in National Meetings<br>**DS151-03:** Larger Data Science Community Interaction |
| | | ***Sub-Category 1.5.2: Individual Teaching/Learning Experience***<br>**DS152-01:** Enrollment in DS Elective Courses<br>**DS152-02:** Statistics PhD<br>**DS152-03:** Math PhD<br>**DS152-04:** Data-intensive PhD<br>**DS152-05:** Master's in applied statistics<br>**DS152-06:** Postdoc in data science education<br>**DS152-07:** Being a tutor/TA in an IDS course<br>**DS152-08:** Self-taught<br>**DS152-09:** Data Science as A Hobby<br>**DS152-10:** Master's in statistics<br>**DS152-11:** Master's in computer science |
| | | ***Sub-Category 1.5.3: Industry Engagement Experience***<br>**DS153-01:** Biostatistician<br>**DS153-02:** Statistical Analyst<br>**DS153-03:** Industry Collaboration |
| | **Category 1.6:** Relationship between experience and teaching DS | **DS16-01:** Interactions and collaborations from DS experts/faculties<br>**DS16-02:** My learning has influenced my teaching |
| **Theme 1:**<br>Data Science (DS)<br>***(Cont'd)*** | | **DS16-03:** Asking good questions to data as a statistician<br>**DS16-04:** Integrating data processing exp. in industry into teaching<br>**DS16-05:** Bringing problem-solving mindset of industry to teaching<br>**DS16-06:** Bringing terminal degree expertise (M.Sc., PhD) to teaching<br>**DS16-07:** Having empathy for students because I didn't get the formal training too.<br>**DS16-08:** Reintroducing teaching experience to improve teaching DS |
| | **Category 1.7:** Change in teaching DS over time | **DS17-01:** No change<br>**DS17-02:** Adding a prerequisite<br>**DS17-03:** No longer co-teach with other departments<br>**DS17-04:** Dividing the initial IDS course into two courses<br>**DS17-05:** Refined pedagogical choices<br>**DS17-06:** Slowing down<br>**DS17-07:** An increase in comfort level in live-coding<br>**DS17-08:** Creating tutorials and lab activities to support coding<br>**DS17-09:** To become more intentional about the data sets that I choose<br>**DS17-10:** Trimming the content<br>**DS17-11:** Changing the content depending on the students' majors and needs.<br>**DS17-12:** Focusing more good programming practices<br>**DS17-13:** Reducing stat & math parts, focusing on programming side<br>**DS17-14:** Adding data science ethics<br>**DS17-15:** Reflecting the evolution of the tools and packages.<br>**DS17-16:** Changing the content depending on the different universities/programs' prioritizations. |
| | **Category 1.8:** Relationship between goals & purposes and teaching DS | **DS18-01:** Teaching data science through data analysis for transferable quantitative skills<br>**DS18-02:** Making the IDS course applied<br>**DS18-03:** Giving them further skills to make them more marketable<br>**DS18-04:** Using real data sets<br>**DS18-05:** Trying to teach them the loop of inquiry<br>**DS18-06:** Informing students what they can/shouldn't do with the data<br>**DS18-07:** Setting up a teaching on data-driven learning |
| | | **DS18-08:** Removing prerequisite to make data science accessible for every student.<br>**DS18-09:** Including critical evaluation of statistical analysis in IDS course design |
| **Theme 2:** Introduction to Data Science (IDS)<br>**Theme 2:** Introduction to Data Science (IDS) (*Cont'd)* | **Category 2.1:** Topics<br><br><br>**Category 2.1:** Topics *(cont'd)* | **IDS221-01:** Data Visualization<br>**IDS221-02:** Data Wrangling<br>**IDS221-03:** Data Science Ethics<br>**IDS221-04:** Simulation<br>**IDS221-05:** Bootstrap<br>**IDS221-06:** Basics of Coding<br>**IDS221-07:** How to Write a Function<br>**IDS221-08:** Modeling<br>**IDS221-09:** *p*-value<br>**IDS221-10:** Data Processing<br>**IDS221-11:** Databases<br>**IDS221-12:** Multivariate Statistics<br>**IDS221-13:** Machine Learning<br>**IDS221-14:** Big Data<br>**IDS221-15:** Web Scraping<br>**IDS221-16:** Data Cleaning<br>**IDS221-17:** Clustering Associations<br>**IDS221-18:** Classification Regression<br>**IDS221-19:** Text Analysis<br>**IDS221-20:** Database Querying<br>**IDS221-21:** Ethics<br>**IDS221-22:** Exploratory Data Analysis<br>**IDS221-23:** File Management<br>**IDS221-24:** Prediction Analysis<br>**IDS221-25:** Linear Regression |
| | | **IDS221-26:** What Is Data Science.<br>**IDS221-27:** What Is Data Scientist<br>**IDS221-28:** KNN<br>**IDS221-29:** Optimization<br>**IDS221-30:** Parameter Estimation<br>**IDS221-31:** Hypothesis Testing<br>**IDS221-32:** Bootstrapping<br>**IDS221-33:** Using A Programming Language<br>**IDS221-34:** Algorithms for Machine Learning |
| **Theme 2:** Introduction to Data Science (IDS) (***Cont'd)*** | **Category 2.2:** Goals of IDS course<br><br><br>**Category 2.2:** Goals of IDS course (***cont'd)*** | **IDS222-01:** To making it relevant to students' daily life<br>**IDS222-02:** To be comfortable with one computer language<br>**IDS222-03:** To teach data ethics<br>**IDS222-04:** To equip with good decision-making skill<br>**IDS222-05:** To ask good questions<br>**IDS222-06:** To build an understanding of the data analytics lifestyle<br>**IDS222-07:** To give an overview of simple statistical models in the basics of machine learning<br>**IDS222-08:** To help students understand the field<br>**IDS222-09:** To grow agency and confidence in the students about developing their ability to perform inquiry in kind of an iterative and connected way<br>**IDS222-10:** To introduce the topic that everybody knows the word but not many people don't know what it means actually<br>**IDS222-11:** To convince technophobic math-phobic student that actually could enjoy, and actually capable of doing this thing.<br>**IDS222-12:** To make students more marketable<br>**IDS222-13:** to give a set of broadly transferable skills that they need to be successful to non-statisticians in other research areas |
| | **Category 2.3:** Interdisciplinarity | **IDS223-01:** Data sets from different sources<br>**IDS223-02:** Students' projects at the end of the semester<br>**IDS223-03:** Draw examples from the life sciences and medicine<br>**IDS223-04:** Try to connect with philosophy<br>**IDS223-05:** COMPAS Recidivism Racial Bias Data |
| | | **IDS223-06:** The course was designed to be interdisciplinary<br>**IDS223-07:** Make sure they get the set of skills that they need to be successful in those other research areas.<br>**IDS223-08:** Primary textbook has interdisc. examples and data sets<br>**IDS223-09:** Classroom examples from different backgrounds<br>**IDS223-10:** Cotaught with librarian to talk about information literacy<br>**IDS223-11:** Guest speaker from economics<br>**IDS223-12:** Project inspired by a biology professor with a fish lab<br>**IDS223-13:** No formal connection<br>**IDS223-14:** data sets from campus offices |
| **Theme 2:** Introduction to Data Science (IDS) (***Cont'd)*** | **Category 2.4:** Technology/Software Used<br><br>**Category 2.4:** Technology/Software Used ***(cont'd)*** | **IDS224-01:** R<br>**IDS224-02:** RStudio<br>**IDS224-03:** Computers<br>**IDS224-04:** PowerPoint<br>**IDS224-05:** RStudio Cloud<br>**IDS224-06:** Faronic Insight<br>**IDS224-07:** Camtasia<br>**IDS224-08:** Clickers<br>**IDS224-09:** SQL<br>**IDS224-10:** GitHub<br>**IDS224-11:** Gradescope<br>**IDS224-12:** Canvas<br>**IDS224-13:** Phyton<br>**IDS224-14:** Jupyter Lab/Notebook<br>**IDS224-15:** Live Code Notebook<br>**IDS224-16:** Tablet/Ipad<br>**IDS224-17:** Google Docs/Slides/Spreadsheets<br>**IDS224-18:** Linux<br>**IDS224-19:** Git<br>**IDS224-20:** GitHub Classroom |
| **Theme 3:** Learning Context | **Category 3.1:** Majors of students | **KOC331-01:** Statistics<br>**KOC331-02:** Mathematics<br>**KOC331-03:** Engineering Students |

| **Theme 3:** Learning Context (*Cont'd)* | | **KOC331-04:** Business School<br>**KOC331-05:** Social Science<br>**KOC331-06:** Accounting Business<br>**KOC331-07:** Economics<br>**KOC 331-08:** Computer Science<br>**KOC331-09:** Physics<br>**KOC331-10:** Humanity<br>**KOC331-11:** Life Sciences<br>**KOC331-12:** Physical Sciences<br>**KOC331-13:** Electrical Engineering<br>**KOC331-14:** Psychology<br>**KOC331-15:** Environmental Science<br>**KOC331-16:** Biology<br>**KOC331-17:** Political Science<br>**KOC331-18:** Data Science<br>**KOC331-19:** Undecided/Undeclared<br>**KOC331-20:** Architecture<br>**KOC331-21:** American Studies<br>**KOC331-22:** Applied Statistics<br>**KOC331-23:** Sociology<br>**KOC331-24:** Cognitive Science<br>**KOC331-25:** English<br>**KOC331-26:** Finance<br>**KOC331-27:** Family Affairs<br>**KOC331-28:** Public Health<br>**KOC331-29:** Ecology<br>**KOC331-30:** Data Analytics<br>**KOC331-31:** Chemistry<br>**KOC331-32:** Wide Variety<br>**KOC331-33:** Philosophy<br>**KOC331-34:** Pre-Med |
|---|---|---|
| | **Category 3.2:** Course capacity | **KOC332-01:** 5-9<br>**KOC332-02:** 10 - 14 |
| **Theme 3:** Learning Context (*Cont'd)* | | **KOC332-03:** 15 - 29<br>**KOC332-04:** 30 – 40<br>**KOC332-05:** 100<br>**KOC332-06:** 130<br>**KOC332-07:** 200<br>**KOC332-08:** 300-450 |
| | **Category 3.3:** Number of Sections | **KOC333-01:** Multiple<br>**KOC333-02:** 1 Section<br>**KOC333-03:** 2-4 Sections<br>**KOC333-04:** 5 Sections |
| | **Category 3.4:** Elective/Required | **KOC334-01:** Both elective and required<br>**KOC334-02:** Elective |
| | **Category 3.5:** Years of Experience | **KOC335-01:** More than years of teaching in IDS<br>**KOC335-02:** Equal to number of years of teaching in IDS |
| | **Category 3.6:** Years of Experience in Teaching IDS | **KOC336-01:** 1<br>**KOC336-02:** 2<br>**KOC336-03:** 3<br>**KOC336-04:** 4<br>**KOC336-06:** 6<br>**KOC336-07:** 7<br>**KOC336-10:** 10<br>**KOC336-11:** Few years |
| | **Category 3.7:** Teaching Team | **KOC337-01:** Solo Instructor<br>**KOC337-02:** Graders<br>**KOC337-03:** Student Mentors<br>**KOC337-04:** Co-instructors<br>**KOC337-05:** More than 1 instructors in multiple sections<br>**KOC337-06:** More than 1 instructors in teaching different quarters/semesters<br>**KOC337-07:** TA(s) |
| | **Category 3.8:** Students' Grade Levels | **KOC338-01:** Freshman<br>**KOC338-02:** Sophomore<br>**KOC338-03:** Junior |

<table>
<tr><td rowspan="7"><b>Theme 3:</b><br>Learning Context<br>(<i><b>Cont'd)</b></i></td><td></td><td><b>KOC338-04:</b> Senior</td></tr>
<tr><td><b>Category 3.9:</b> Department Offered IDS</td><td><b>KOC339-01:</b> Mathematics and statistics<br><b>KOC339-02:</b> Mathematics<br><b>KOC339-03:</b> Statistics<br><b>KOC339-04:</b> Center/Institute<br><b>KOC339-05:</b> Statistics and Data Science<br><b>KOC339-06:</b> Other<br><b>KOC339-07:</b> Mathematics and Computer Science<br><b>KOC339-08:</b> Data Analytics</td></tr>
<tr><td><b>Category 3.10:</b> University Type</td><td><b>KOC3310-01:</b> Research University<br><b>KOC3310-02:</b> Liberal Arts College</td></tr>
<tr><td><b>Category 3.11:</b> Semester/Quarter</td><td><b>KOC3311-01:</b> Semester<br><b>KOC3311-02:</b> Quarter<br><b>KOC3311-03:</b> Other</td></tr>
<tr><td><b>Category 3.12:</b> Terminal Degree of IDS Instructors</td><td><b>KOC3312-01:</b> Biostatistics PhD<br><b>KOC3312-02:</b> Mathematics PhD<br><b>KOC3312-03:</b> Statistics PhD<br><b>KOC3312-04:</b> Statistics Master's Degree<br><b>KOC3312-05:</b> Computer Science Master's Degree<br><b>KOC3312-06:</b> Statistics and Data Analytics PhD<br><b>KOC3312-07:</b> Other (Genetics, Economics)<br><b>KOC3312-08:</b> Math Education PhD</td></tr>
<tr><td><b>Category 3.13:</b> Formal Training in DS</td><td><b>KOC3313-01:</b> Self-taught<br><b>KOC3313-02:</b> DS-Related Terminal Degree Expertise<br><b>KOC3313-03:</b> Workshops<br><b>KOC3313-04:</b> Industry Experience<br><b>KOC3313-05:</b> DS-Related Course Enrollment In Graduate Years</td></tr>
<tr><td><b>Category 3.14:</b> Formal Training in Pedagogy/Teaching</td><td><b>KOC3314-01:</b> No terminal degree<br><b>KOC3314-02:</b> Workshops<br><b>KOC3314-03:</b> Conducting Pedagogical Research<br><b>KOC3314-05:</b> Pedagogical Course Enrollment In (under)graduate Years<br><b>KOC3314-06:</b> Terminal Degree/Post-doc in Pedagogy</td></tr>
</table>

| | | **KOC3314-07:** Teaching Internship in PhD<br>**KOC3314-08:** TA Trainings<br>**KOC3314-09:** Self-taught |
|---|---|---|
| **Theme 4:** Orientation to teach IDS | **Category 4.1:** Enhance Data Literacy/ Teach to Learn from Data | **OTIDS441-01:** Reinforcing Statistical Thinking for CS Students<br>**OTIDS441-02:** Expanding Data Engagement Beyond CS and Statistics<br>**OTIDS441-03:** Enhancing Understanding of the Field<br>**OTIDS441-04:** Promoting Agency and Confidence in Inquiry<br>**OTIDS441-05:** Familiarizing Students with DS Tools and Concepts<br>**OTIDS441-06:** Developing Appreciation for Data Science Ubiquity<br>**OTIDS441-07:** Equipping Students with Essential Data Science Skills for Diverse Majors/Research Areas |
| **Theme 4:** Orientation to teach IDS ***(Cont'd)*** | **Category 4.2:** Familiarize Students with A Programming Language | **OTIDS442-01:** Fostering Programming Skills for Statistics Students<br>**OTIDS442-02:** Promoting Proficiency in R |
| | **Category 4.3:** Attracting Students to Major/Minor in Data Science | **OTIDS443-01:** Introducing Students What Data Science is.<br>**OTIDS443-02:** Generating Interest in Further DS Courses<br>**OTIDS443-03:** Increasing Recruitment for Data Science Major<br>**OTIDS443-04:** Giving them further skills to make them more marketable.<br>**OTIDS443-05:** Cultivating Student Enthusiasm for Data Science (Major) |
| **Theme 5:** Knowledge of Students' Understanding | **Category 5.1:** Prior Knowledge | **KOSIDSSYN551-01:** Algebraic Understanding<br>**KOSIDSSYN551-02:** Understanding of Functions<br>**KOSIDSSYN551-03:** Visualizing Algebraic Concepts<br>**KOSIDSSYN551-04:** Coding Proficiency (As a suggestion)<br>**KOSIDSSYN551-05:** Comfort with Computers<br>**KOCIDSSYN551-06:** College-level Writing Skills<br>**KOCIDSSYN551-07:** No Prior Knowledge<br>**KOCIDSSYN551-08:** Basic Programming Background<br>**KOCIDSSYN551-09:** Understanding of Derivatives<br>**KOCIDSSYN551-10:** Basic Analytical Skills |

<table>
<tr>
<td rowspan="5"><b>Theme 5:</b> Knowledge of Students' Understanding (<b><i>Cont'd)</i></b></td>
<td colspan="2"></td>
<td><b>KOCIDSSYN551-11:</b> High School Algebra Proficiency<br><b>KOCIDSSYN551-12:</b> Basic Math</td>
</tr>
<tr>
<td rowspan="2"><b>Category 5.2:</b> Learning Difficulties</td>
<td rowspan="2"><b>Category 5.2.1:</b> Syntactic Knowledge Difficulties</td>
<td><b><i>Sub-Category 5.2.1.1: Markup Languages and Reproducibility Tools</i></b><br><b>KOSIDSSYN55211-01:</b> HTML<br><b>KOSIDSSYN55211-02:</b> Jupyter Notebook<br><b>KOSIDSSYN55211-03:</b> R Markdown<br><b>KOSIDSSYN55211-04:</b> Linux<br><b>KOSIDSSYN55211-05:</b> Quarto Markdown<br><b>KOSIDSSYN55211-06:</b> Git/GitHub</td>
</tr>
<tr>
<td><b><i>Sub-Category 5.2.1.2: Programming Languages</i></b><br><b>KOSIDSSYN55212-01:</b> Packages<br><b>KOSIDSSYN55212-02:</b> Adapting the Code<br><b>KOSIDSSYN55212-03:</b> Libraries<br><b>KOSIDSSYN55212-04:</b> How to Read Data<br><b>KOSIDSSYN55212-05:</b> Misspelling</td>
</tr>
<tr>
<td rowspan="2"><b>Category 5.2:</b> Learning Difficulties <b><i>(cont'd)</i></b></td>
<td rowspan="2"><b>Category 5.2.2:</b> Conceptual Knowledge Difficulties</td>
<td><b><i>Sub-Category 5.2.2.1: Mathematics_Difficulties</i></b><br><b>KOSIDSCON55221-01:</b> Algorithms<br><b>KOSIDSCON55221-02:</b> Permutation Testing</td>
</tr>
<tr>
<td><b><i>Sub-Category 5.2.2.2: Statistics_Difficulties</i></b><br><b>KOSIDSCON55222-01:</b> Types of Variables<br><b>KOSIDSCON55222-02:</b> Confidence Interval<br><b>KOSIDSCON55222-03:</b> Principles of Data Viz.<br><b>KOSIDSCON55222-04:</b> Hypothesis Testing<br><b>KOSIDSCON55222-05:</b> Correlation vs. Causality<br><b>KOSIDSCON55222-06:</b> Bootstrapping<br><b>KOSIDSCON55222-07:</b> Inductive Inference<br><b>KOSIDSCON55222-08:</b> Statistical Analysis Methods-Modelling<br><b>KOSIDSCON55222-09:</b> p-value<br><b>KOSIDSCON55222-10:</b> Sampling Distribution</td>
</tr>
</table>

<table>
<tr>
<td rowspan="4"><b>Theme 5:</b> Knowledge of Students' Understanding (<i><b>Cont'd)</b></i></td>
<td rowspan="4"><b>Category 5.2:</b> Learning Difficulties <i><b>(cont'd)</b></i></td>
<td rowspan="3"></td>
<td><i><b>Sub-Category 5.2.2.3: Computer Science_Difficulties</b></i><br><b>KOSIDSCON55223-01:</b> I/O File Management<br><b>KOSIDSCON55223-02:</b> Working Mech. of Markup Languages<br><b>KOSIDSCON55223-03:</b> Basics of Coding<br><b>KOSIDSCON55223-04:</b> Filter Function<br><b>KOSIDSCON55223-05:</b> Basics of Web Scraping<br><b>KOSIDSCON55223-06:</b> Select Function<br><b>KOSIDSCON55223-07:</b> Joining Data Sets<br><b>KOSIDSCON55223-08:</b> Mapping Functions<br><b>KOSIDSCON55223-09:</b> Loops<br><b>KOSIDSCON55223-10:</b> Creating Functions</td>
</tr>
<tr>
<td><i><b>Sub-Category 5.2.2.4: Domain – Specific Knowledge_Difficulties</b></i><br><b>KOSIDSCON55224-01:</b> Understanding the Nature of Data<br><b>KOSIDSCON55224-02:</b> Understanding Technical Writing</td>
</tr>
<tr>
<td><i><b>Sub-Category 5.2.2.5: Interdisciplinary Knowledge_Difficulties</b></i><br><b>KOSIDSCON55225-01:</b> Ethics<br><b>KOSIDSCON55225-02:</b> Machine Learning</td>
</tr>
<tr>
<td><b>Category 5.2.3:</b> Strategic Knowledge</td>
<td><b>KOSIDSSTR5523-01:</b> Debugging<br><b>KOSIDSSTR5523-02:</b> Communication<br><b>KOSIDSSTR5523-03:</b> Data Wrangling<br><b>KOSIDSSTR5523-04:</b> Appreciating the complexity of Interdisciplinary Research<br><b>KOSIDSSTR5523-05:</b> Making Appropriate Data Visualization Decisions<br><b>KOSIDSSTR5523-06:</b> Creative Thinking<br><b>KOSIDSSTR5523-07:</b> Proper Use of Descriptive Statistics<br><b>KOSIDSSTR5523-08:</b> Conducting a Good Research<br><b>KOSIDSSTR5523-09:</b> Deciding Statistical Analysis Methods-Modelling<br><b>KOSIDSSTR5523-10:</b> Working with Real and Messy Data<br><b>KOSIDSSTR5523-11:</b> Handling Missing Data<br><b>KOSIDSSTR5523-12:</b> Asking Good Questions<br><b>KOSIDSSTR5523-13:</b> Web Scraping</td>
</tr>
</table>

| | | | **KOSIDSSTR5523-14:** Setting up Data Science Pipeline |
|---|---|---|---|
| | | **Category 5.2.4:** Adjustment of teaching | **KOSIDSADJ5524-01:** Support struggling students during class<br>**KOSIDSADJ5524-02:** Students' feedbacks<br>**KOSIDSADJ5524-03:** Create collaborative learning environment<br>**KOSIDSADJ5524-04:** Providing a balance between lecturing and learning by doing it<br>**KOSIDSADJ5524-05:** Creating a curriculum accessible to a wide range of students<br>**KOSIDSADJ5524-06:** Limiting the number of concepts (1-2 related concepts) in each lecture |
| **Theme 5:** Knowledge of Students' Understanding (***Cont'd)*** | **Category 5.3:** Misconceptions/Stereotypes<br><br>**Category 5.3:** Misconceptions/Stereotypes ***(cont'd)*** | **Category 5.3.1:** Statements | **KOSIDSMIS5531-01:** Defining data science as synonymous with statistics.<br>**KOSIDSMIS5531-02:** It's easy to find the data that you want.<br>**KOSIDSMIS5531-03:** If you can code quickly, you're a good data scientist.<br>**KOSIDSMIS5531-04:** Men are better at coding, they are faster, more efficient, and quicker than females.<br>**KOSIDSMIS5531-05:** If I do DS instead of business analytics, when I get out, I can get a well-paying job.<br>**KOSIDSMIS5531-06:** Data scientists are tech-bros<br>**KOSIDSMIS5531-07:** Students don't see the cyclic nature of DS<br>**KOSIDSMIS5531-08:** They need to be a really phenomenal programmer when they walk in the door<br>**KOSIDSMIS5531-09:** They're either good or bad at data science (lack of growth mindset).<br>**KOSIDSMIS5531-10:** If they have strong programming backgrounds, they think they're going to be great.<br>**KOSIDSMIS5531-11:** Data science is machine learning.<br>**KOSIDSMIS5531-12:** Some students don't come in with any internal distinctions between data science, machine learning, artificial intelligence. |

<table>
<tr>
<td rowspan="4"><b>Theme 5:</b><br>Knowledge of Students' Understanding (<b><i>Cont'd)</i></b></td>
<td rowspan="2"><b>Category 5.3:</b><br>Misconceptions/Stereotypes<br><b><i>(cont'd</i></b></td>
<td><b>Category 5.3.1:</b><br>Statements<br><b><i>(cont'd)</i></b></td>
<td><b>KOSIDSMIS5531-13:</b> Data science is a field to be a collection of big hammers that if they learn how to swing each of these hammers, they could whack every problem in the world.<br><b>KOSIDSMIS5531-14:</b> If a professional has analyzed the data, then that is now trustworthy.<br><b>KOSIDSMIS5531-15:</b> data scientist is a person who is really good at numbers, analyze numbers like crazy, and spend nights working from a computer analyzing data<br><b>KOSIDSMIS5531-16:</b> They thought that they didn't look like a data scientist<br><b>KOSIDSMIS5531-17:</b> They are hesitant to trust data science</td>
</tr>
<tr>
<td><b>Category 5.3.2:</b><br>Sources</td>
<td><b>KOSIDSMIS5532-01:</b> Seeing end products in data science creates an outcome-centric perception bias.<br><b>KOSIDSMIS5532-02:</b> Society<br><b>KOSIDSMIS5532-03:</b> Popular Culture<br><b>KOSIDSMIS5532-04:</b> Cultural Biases<br><b>KOSIDSMIS5532-05:</b> Lack of Knowledge<br><b>KOSIDSMIS5532-06:</b> White and Male dominant<br><b>KOSIDSMIS5532-07:</b> Industry<br><b>KOSIDSMIS5532-08:</b> Media<br><b>KOSIDSMIS5532-09:</b> Peers<br><b>KOSIDSMIS5532-10:</b> YouTube<br><b>KOSIDSMIS5532-11:</b> Data from Star Trek<br><b>KOSIDSMIS5532-12:</b> Demographics in the major<br><b>KOSIDSMIS5532-13:</b> I have no answer for that</td>
</tr>
<tr>
<td colspan="2" rowspan="2"><b>Category 5.4:</b> Motivation & interest</td>
<td><b><i>Sub-Category 5.4.1: Motivation & Encouragement</i></b><br><b>KOSIDSMIS5541-01:</b> Keeping them motivated and optimistic is important.<br><b>KOSIDSMIS5541-02:</b> Nudging them for writing about the procedures important.</td>
</tr>
<tr>
<td><b><i>Sub-Category 5.4.2: Frustration and Challenges</i></b><br><b>KOSIDSMIS5542-01:</b> They are frustrated that it's not a simple task, very similar to math anxiety.</td>
</tr>
</table>

| **Theme 5:** Knowledge of Students' Understanding (***Cont'd)*** | | **KOSIDSMIS5542-02:** At first, they don't understand that programming is trial and error a lot. So, they get frustrated.<br>**KOSIDSMIS5542-03:** Their motivation decreases when they see error message.<br>**KOSIDSMIS5542-04:** Sometimes they are frustrated when they failed to complete a task in R |
|---|---|---|
| | | ***Sub-Category 5.4.3: Self-Efficacy and Computer Literacy***<br>**KOSIDSMIS5543-01:** Computers are smarter than me.<br>**KOSIDSMIS5543-02:** My computer hates me.<br>**KOSIDSMIS5543-03:** R hates me.<br>**KOSIDSMIS5543-04:** My R is broken.<br>**KOSIDSMIS5543-05:** I am just really bad at computers.<br>**KOSIDSMIS5543-06:** I am not very good at math dynamic. |
| | | ***Sub-Category 5.4.4: Student Resistance***<br>**KOSIDSMIS5544-01:** Every time I do something mathy, the computer scientists get grumpy.<br>**KOSIDSMIS5544-02:** Anytime I need to teach a math concept is generally when my students give me the most resistance.<br>**KOSIDSMIS5544-03:** Math and computer science students have a higher motivation then their peers. |
| | **Category 5.5:** Extra Support | **KOSIDSMIS5551-01:** Support struggling students during office hours.<br>**KOSIDSMIS5551-02:** Extra sessions<br>**KOSIDSMIS5551-03:** Prerecorded R tutorials<br>**KOSIDSMIS5551-04:** Student mentors in class<br>**KOSIDSMIS5551-05:** Trying to break it down into small pieces.<br>**KOSIDSMIS5551-06:** Trying to explain something in a different way than you explained it before.<br>**KOSIDSMIS5551-07:** Developing math tutorials for programmers<br>**KOSIDSMIS5551-08:** Reviewing Slides before starting the lecture<br>**KOSIDSMIS5551-09:** Acknowledging it is okay to be bad at coding at the beginning<br>**KOSIDSMIS5551-10:** Being a Prompt Respondent to Student Emails |

| | | | **KOSIDSMIS5551-11:** Helping them occasionally |
|---|---|---|---|
| **Theme 6:** Knowledge of Instructional Strategies | **Category 6.1:** Subject Specific | | **KOISIDS661-01:** Flipped Classroom<br>**KOISIDS661-02:** Lecturing enhanced with reading, videos, interactive visualization<br>**KOISIDS661-03:** Discussion<br>**KOISIDS661-04:** Live Coding<br>**KOISIDS661-05:** Project-Based Learning<br>**KOISIDS661-06:** Lecturing<br>**KOISIDS661-07:** Questioning<br>**KOISIDS661-08:** Demonstration<br>**KOISIDS661-09:** Group Work<br>**KOISIDS661-10:** Think-Pair-Share<br>**KOISIDS661-11:** Interactive Learning<br>**KOISIDS661-12:** Collaborative Learning<br>**KOISIDS661-13:** Simulation based learning<br>**KOISIDS661-14:** Active learning<br>**KOISIDS661-15:** Guided practice |
| | **Category 6.2:** Topic specific | | **KOISIDS662-01:** Storytelling (real-world cases)<br>**KOISIDS662-02:** Questioning (data ethics)<br>**KOISIDS662-03:** Role Playing (joining data sets)<br>**KOISIDS662-04:** Tactile simulation (sampling distribution) |
| **Theme 7:** Knowledge of Curriculum | **Category 7.1:** Vertical Curriculum Alignment | **Category 7.1.1:** Prerequisite | **KOCIDS7711-01:** AP statistics<br>**KOCIDS7711-02:** College-level statistics<br>**KOCIDS7711-03:** Programming experience (recommended)<br>**KOCIDS7711-04:** Introductory programming (required)<br>**KOCIDS7711-05:** Introductory Statistics<br>**KOCIDS7711-06:** None<br>**KOCIDS7711-07:** Corequisite with a computing class<br>**KOCIDS7711-08:** Introductory Computer Science<br>**KOCIDS7711-09:** Single Variable Calculus<br>**KOCIDS7711-10:** Basic Math<br>**KOCIDS7711-11:** Introductory Statistics (recommended) |

<table>
<tr>
<td rowspan="4"><b>Theme 7:</b><br>Knowledge of Curriculum<br>(<b><i>Cont’d)</i></b></td>
<td></td>
<td><b>Category 7.1.2:</b> Prerequisite to any other</td>
<td><b>KOCIDS7712-01:</b> None<br><b>KOCIDS7712-02:</b> Machine Learning<br><b>KOCIDS7712-03:</b> More advanced DS courses<br><b>KOCIDS7712-04:</b> Senior Capstone<br><b>KOCIDS7712-05:</b> Data Journalism<br><b>KOCIDS7712-06:</b> Advance Programming for DS<br><b>KOCIDS7712-07:</b> Statistical Learning<br><b>KOCIDS7712-08:</b> Capstone courses<br><b>KOCIDS7712-09:</b> All others in the data analytics program<br><b>KOCIDS7712-10:</b> Not a requisite but recommended class<br><b>KOCIDS7712-11:</b> Data Acquisition<br><b>KOCIDS7712-12:</b> Big Data<br><b>KOCIDS7712-13:</b> Yes, but they don’t know the name</td>
</tr>
<tr>
<td><b>Category 7.2:</b> Curricular Saliency</td>
<td><b>Category 7.2.1:</b> Effects of Prereq. to Sequencing</td>
<td><b>KOCIDS7721-01:</b> No Effect<br><b>KOCIDS7721-02:</b> Yes<br><b>KOCIDS7721-03:</b> adding GitHub<br><b>KOCIDS7721-04:</b> Increasing the amount of statistical thinking/inference<br><b>KOCIDS7721-05:</b> spend more time teaching background because can't assume more prerequisites (e.g., calculus, linear algebra, data structure, statistics)<br><b>KOCIDS7721-06:</b> Do not know</td>
</tr>
<tr>
<td colspan="2"><b>Category 7.3:</b> Curricular Materials</td>
<td><b>KOCIDS773-01:</b> Self-developed<br><b>KOCIDS773-02:</b> Utilizing textbook resources<br><b>KOCIDS773-03:</b> Drawn from comparable online IDS courses<br><b>KOCIDS773-04:</b> Collaboratively crafted by colleagues<br><b>KOCIDS773-05:</b> Revised with input from student feedback<br><b>KOCIDS773-06:</b> Used a coworker’s structure as a guide</td>
</tr>
<tr>
<td colspan="2"><b>Category 7.4:</b> Curricular Guideline</td>
<td><b>KOCIDS774-01:</b> ACM guidelines<br><b>KOCIDS774-02:</b> ASA guidelines<br><b>KOCIDS774-03:</b> No guideline<br><b>KOCIDS774-04:</b> Park City Institute guideline</td>
</tr>
</table>

| | | | |
|---|---|---|---|
| | | | **KOCIDS774-05:** Developed by an interdisciplinary team<br>**KOCIDS774-06:** GAISE recommendations<br>**KOCIDS774-07:** Department goals<br>**KOCIDS774-08:** Guidelines developed by other stat ed people |
| **Theme 8:** Knowledge of Assessment | **Category 8.1:** Reasons for Assessment | | **KOAIDS881-01:** to inform my own teaching<br>**KOCIDS881-02:** to keep the pace of learning<br>**KOCIDS881-03:** to monitor students' progress<br>**KOCIDS881-04:** to catch the difficulties or misconceptions with the material early<br>**KOCIDS881-05:** to make my job easier as a grader<br>**KOCIDS881-06:** to get them to focus on what I really wanted them to.<br>**KOCIDS881-07:** encouraging students to interact with material.<br>**KOCIDS881-08:** Institutional requirement (summative assessment)<br>**KOCIDS881-09:** To give feedback (formative)<br>**KOCIDS881-10:** to test whether or not a student was able to learn all the things that we're taught in class<br>**KOCIDS881-11:** encouraging students to attend the classes by administering clicker quizzes<br>**KOCIDS881-12:** to develop an intuition for what to expect<br>**KOCIDS881-13:** Using ungrading to create more equitable, inclusive and less anxious learning environments |
| **Theme 8:** Knowledge of Assessment (***Cont'd)*** | **Category 8.2:** Types of Assessment | **Category 8.2.1:** Summative | **KOAIDS8821-01:** Peer evaluation<br>**KOAIDS8821-02:** (Group) Projects<br>**KOAIDS8821-03:** Weekly Homework<br>**KOAIDS8821-04:** Reading Assignment<br>**KOAIDS8821-05:** Quiz<br>**KOAIDS8821-06:** Online Quiz<br>**KOAIDS8821-07:** Exam<br>**KOAIDS8821-08:** Auto-graded assessment<br>**KOAIDS8821-09:** Biweekly Homework<br>**KOAIDS8821-10:** Lab assignments |
| | | | **KOAIDS8821-11:** Take home exam<br>**KOAIDS8821-12:** Classroom discussions |
| | | **Category 8.2.2:** Formative | **KOAIDS8822-01:** Selecting a student to code together<br>**KOAIDS8822-02:** Selecting a student to ask a question<br>**KOAIDS8822-03:** Informal assessment during the class<br>**KOAIDS8822-04:** In-class assessment & feedback<br>**KOAIDS8822-05:** Providing detailed feedback outside of the class |
| | | **Category 8.2.3:** Ungrading | **KOAIDS8823-01:** more equitable learning environments<br>**KOAIDS8823-02:** more inclusive learning environments<br>**KOAIDS8823-03:** less stress and anxiety for students |
| | **Category 8.3:** Types of Feedback | | **KOAIDS883-01:** Handwritten<br>**KOAIDS883-02:** Oral<br>**KOAIDS883-03:** Both written and oral<br>**KOAIDS883-04:** Auto-graded feedback<br>**KOAIDS883-05:** A list of commonly wrong answers |
| | **Category 8.4:** Taxonomy Used | | **KOAIDS884-01:** Bloom's Taxonomy |
| **Theme 9:** Self-Efficacy Beliefs<br><br>**Theme 9:** Self-Efficacy Beliefs (***Cont'd)*** | **Category 9.1:** Content | **Category 9.1.1:** More comfortable | **SEFFBIDS9911-01:** Data wrangling<br>**SEFFBIDS9911-02:** Data visualization<br>**SEFFBIDS9911-03:** Ethics<br>**SEFFBIDS9911-04:** Tidy data<br>**SEFFBIDS9911-05:** Data science ethics<br>**SEFFBIDS9911-06:** using R<br>**SEFFBIDS9911-07:** Producing effective data visuals<br>**SEFFBIDS9911-08:** Asking good questions about the data<br>**SEFFBIDS9911-09:** Making good decisions about the data<br>**SEFFBIDS9911-10:** All the topics in IDS<br>**SEFFBIDS9911-11:** Statistics side topics<br>**SEFFBIDS9911-12:** SQL<br>**SEFFBIDS9911-13:** Debugging<br>**SEFFBIDS9911-14:** Web scraping<br>**SEFFBIDS9911-15:** How to teach programming<br>**SEFFBIDS9911-16:** Mathematical and algorithmic concepts because of my training |
| | | | **SEFFBIDS9911-17:** Modeling<br>**SEFFBIDS9911-18:** Clustering<br>**SEFFBIDS9911-19:** Always find a solution to the problems in my class<br>**SEFFBIDS9911-20:** Critical thinking<br>**SEFFBIDS9911-21:** Storytelling with data |
| | | **Category 9.1.2:** Less Comfortable | **SEFFBIDS9912-01:** Computer science side of DS<br>**SEFFBIDS9912-02:** Machine learning<br>**SEFFBIDS9912-03:** Shiny applications<br>**SEFFBIDS9912-04:** Big Data Management<br>**SEFFBIDS9912-05:** Hadoop<br>**SEFFBIDS9912-06:** Geospatial data<br>**SEFFBIDS9912-07:** Ethics<br>**SEFFBIDS9912-08:** Tidy Models<br>**SEFFBIDS9912-09:** Mapping functions<br>**SEFFBIDS9912-10:** Decision trees<br>**SEFFBIDS9912-11:** Random forests<br>**SEFFBIDS9912-12:** Text analysis<br>**SEFFBIDS9912-13:** Databases |
| | | **Category 9.1.3:** Want to know more | **SEFFBIDS9913-01:** HTML<br>**SEFFBIDS9913-02:** Ethics<br>**SEFFBIDS9913-03:** Computer science side of DS |
| | | **Category 9.1.4:** Middle Comfort | **SEFFBIDS9914-01:** All the topics in IDS (middle confident)<br>**SEFFBIDS9914-02:** Optimization<br>**SEFFBIDS9914-03:** Indexes<br>**SEFFBIDS9914-04:** HTML |
| **Theme 9:** Self-Efficacy Beliefs | **Category 9.2:** Pedagogy | **Category 9.2.1:** More comfortable | **SEFFBIDS9921-01:** Socratic Method<br>**SEFFBIDS9921-02:** Designing Assignments<br>**SEFFBIDS9921-03:** Helping students find the errors in their code.<br>**SEFFBIDS9921-04:** Lecturing<br>**SEFFBIDS9921-05:** Answering questions during my office hours<br>**SEFFBIDS9921-06:** Live coding<br>**SEFFBIDS9921-07:** Assessment<br>**SEFFBIDS9921-08:** Answering questions during the class time |
| | | **Category 9.2.2:** Less Comfortable | **SEFFBIDS9922-01:** Lecturing as a method<br>**SEFFBIDS9922-02:** How to design a course<br>**SEFFBIDS9922-03:** Hypothesis testing<br>**SEFFBIDS9922-04:** Permutation testing<br>**SEFFBIDS9922-05:** Not easy while teaching sometimes due to being a new faculty<br>**SEFFBIDS9922-06:** How to teach coding |
| | | **Category 9.2.3:** Want to know more | **SEFFBIDS9923-01:** what other skills the students need to achieve gainful employment in a data science job afterwards.<br>**SEFFBIDS9923-02:** to leverage the students' strengths and weaknesses, or leverage the strength that they bring into the classroom. |